\documentclass{jfm}
\usepackage{graphicx}
\usepackage{subfig}
\usepackage{epstopdf, epsfig}
\usepackage{color}
\usepackage[dvipsnames]{xcolor}

\graphicspath{{./fig_exp/},{./fig_sim/}}

\title{Single cavitation bubble dynamics in a stagnation flow}
 
\author{Dominik Mnich\aff{1},
  Fabian Reuter\aff{1},\\
  Fabian Denner\aff{2}
 \and Claus-Dieter Ohl\aff{1}}

\affiliation{
\aff{1} Faculty of Natural Sciences, Institute for Physics, Otto-von-Guericke-University Magdeburg, Universitätsplatz 2, 39106 Magdeburg, Germany
\aff{2} Department of Mechanical Engineering, Polytechnique Montr\'eal, Montr\'eal, H3T 1J4, QC, Canada
}

\begin{document}
\newcounter{Picture}

\maketitle

\begin{abstract}
Jetting of collapsing bubbles is a key aspect in cavitation-driven fluid-solid interactions as it shapes the bubble dynamics and additionally due to its direct interaction with the wall. We study experimentally and numerically the near-wall collapse and jetting of a single bubble seeded into the stagnation flow of a wall jet, i.e. a jet that impinges perpendicular onto a solid wall. High-speed imaging shows rich and rather distinct bubble dynamics for different wall jet flow velocities and bubble-to-wall stand-off distances. The simulations use a Volume-of-Fluid method and allow to numerically determine the microscopic and transient pressures and shear stresses on the wall. It is shown that a wall jet at moderate flow velocities of a few metres per second already shapes the bubble ellipsoidally inducing a planar and convergent jet flow. The distinct bubble dynamics allow to tailor the wall interaction. In particular, the shear stresses can be increased by orders of magnitude without increasing impact pressures the same way. Interestingly, at small seeding stand-offs, the bubble during the final collapse stage can lift off the wall and migrate against the flow direction of the wall jet such that the violent collapse occurs away from the wall.
\end{abstract}

\begin{keywords}

\end{keywords}

\section{Introduction}

In the vicinity of a solid boundary, cavitation bubbles in stagnant liquids collapse non-spherically. They result in jetting and shockwave emissions and expose the boundary to significant wall shear stresses and pressures. The dynamics of the simple case of a bubble collapsing near a rigid planar surface has been the subject of many studies in the last decades, see for example \citep{plesset_chapman_1971,lauterborn_bolle_1975,zhang_duncan_chahine_1993,Supponen2016}. Jets pierce the bubble and are typically directed toward the rigid boundary. They can reach velocities of $50-100~$ms$^{-1}$\citep{Ellis1966,Blake1987,Lauterborn2010}, for millimeter-sized bubbles at atmospheric pressure, known as regular jetting. The most important parameter that determines the bubble dynamics in a stagnant liquid and in the absence of gravity is the non-dimensional stand-off distance $\gamma=d/R_\mathrm{max}$, where $d$ is the initial distance of bubble nucleation from the wall and $R_\mathrm{max}$ is the radius of the bubble at its maximum expansion. For the smallest bubble-to-wall stand-off distances, even velocities of more than one order of magnitude larger are possible, known as needle jetting \citep{Lechner2019,Reuter2021,Bussmann2023}. Most violent collapses, i.e. erosive cavitation, occur when a bubble during its collapse focuses the self-emitted shockwaves onto itself to produce a shockwave self-intensified collapse. This was observed for bubbles of very small stand-off distances, i.e., $\gamma \lesssim 0.2$ by \citet{reuter2022cavitation}.

A particularly interesting case is when the bubble takes the shape of a slender spherical cap during maximum expansion, which leads to a liquid jet that is directed away from the boundary. This kind of jet has been predicted for mildly ellipsoidal capped bubbles by \citet{Lauer2012} and \citet{aganin2019dynamics}, and was later shown to occur for axisymmetric cap-shaped bubbles in simulations and experiments by \citet{Saini2022}. Here, a high-pressure region below the bubble forms that accelerates a liquid jet flow away from the surface. Other configurations involving elastic boundaries \citep{D.C.Gibson1966,Brujan2001} or a narrow gap \citep{Chahine1982,Zeng2020,Gonzalez-Avila2011b} result in hourglass-shaped bubbles that pinch off and give rise to two fast and rather thin jets in opposite directions. 

The broad range of bubble dynamics allows a wide field of useful applications, such as surface cleaning \citep{Ohl2006,Kim2009,Reuter2016,Gonzalez-Avila2011,Reuter2017membrane,Ando2021, Ando2023} or drug delivery via sonoporation \citep{Prentice2005,Gac2007,Ohl2006Sono}. 
%Long-lived vortices can help to exhibit moderate shear stresses over larger areas on the solid.
For these applications, large wall shear stresses are beneficial, especially if they can be generated without exposing the wall to an excessively large pressure that may cause damage to even the hardest materials \citep{Tomita1986,Philipp1998,reuter2022cavitation}. 
Wall shear stresses from a single bubble collapse have been measured \citep{dijkink2008measurement,reuter2018electrochemical} and simulated \citep{koukouvinis2018parametric,zeng2022wall} to reach peak values of $100\,$kPa. Experimental data yield peak values that are one order of magnitude smaller, which is still compatible, as sampling periods were one order of magnitude larger than the shear spike durations.

While these observations are valid for bubbles in an initially stagnant liquid, the presence of a flow can change the bubble dynamics and, thus, the wall shear stress and pressure generated at the wall drastically.
One interesting case is hydrodynamic cavitation in a stagnation flow, as it was shown by \citet{knapp1955recent} that the strongest erosion for a model in a water tunnel can be found in the area around stagnation points.
So far, research on the effect of a flow on the dynamics of a {\em single} bubble was conducted through simulations by \citet{blake_taib_doherty_1986}, \citet{robinsonblake1994}
 and \citet{Blake2015}. 
 %dabei aber weder kleine abstände, viskosität (nicht in BEM) und keine untersuchung der wall shear stress
 They utilised a boundary element method to study the influence of an ideal stagnation point flow on the gross bubble dynamics and, in particular, on the bubble's jetting behaviour. They found hourglass-shaped bubbles that pinch off, resulting in much faster and thinner jets. Experimental research on the dynamics of bubbles in flows near a stagnation point is rather scarce. \citet{Starrett1982} reported in his thesis of a striking contrast between the bubble dynamics in a quiescent liquid and in a stagnation flow. He observed that during expansion the bubble deforms into an ellipsoid with the major axis aligned parallel to the wall, then the bubble forms a waist at its equator that divides the bubble into two parts just prior to collapse. Starret studied the dynamics of rather large bubbles created with an electric discharge in stagnation flows with free stream velocities between about $6 \mbox{--} 18\,$m/s and concentrated on larger distances rather than on small stand-offs, which are more relevant for the bubble-wall interaction. 

In the present work, we study the dynamics of a single cavitation bubble in a wall jet in particular at rather low flow velocities and small stand-off distances, i.e., where viscosity plays an important role. This regime is relevant for many applications. In experiments, we record the bubble dynamics with a suitable high-speed camera to resolve the details of the collapse, whereas the simulations, which include the effects of compressibility and viscosity, allow determining the wall shear stresses and pressure acting on the wall as a consequence of the bubble collapse. The results reveal that a bubble collapse with beneficial high wall shear stresses can be realised while avoiding potentially damaging behaviour.

\section{Methods}

\subsection{Experimental methods}

\begin{figure}
\centering
\includegraphics[width=10.5cm]{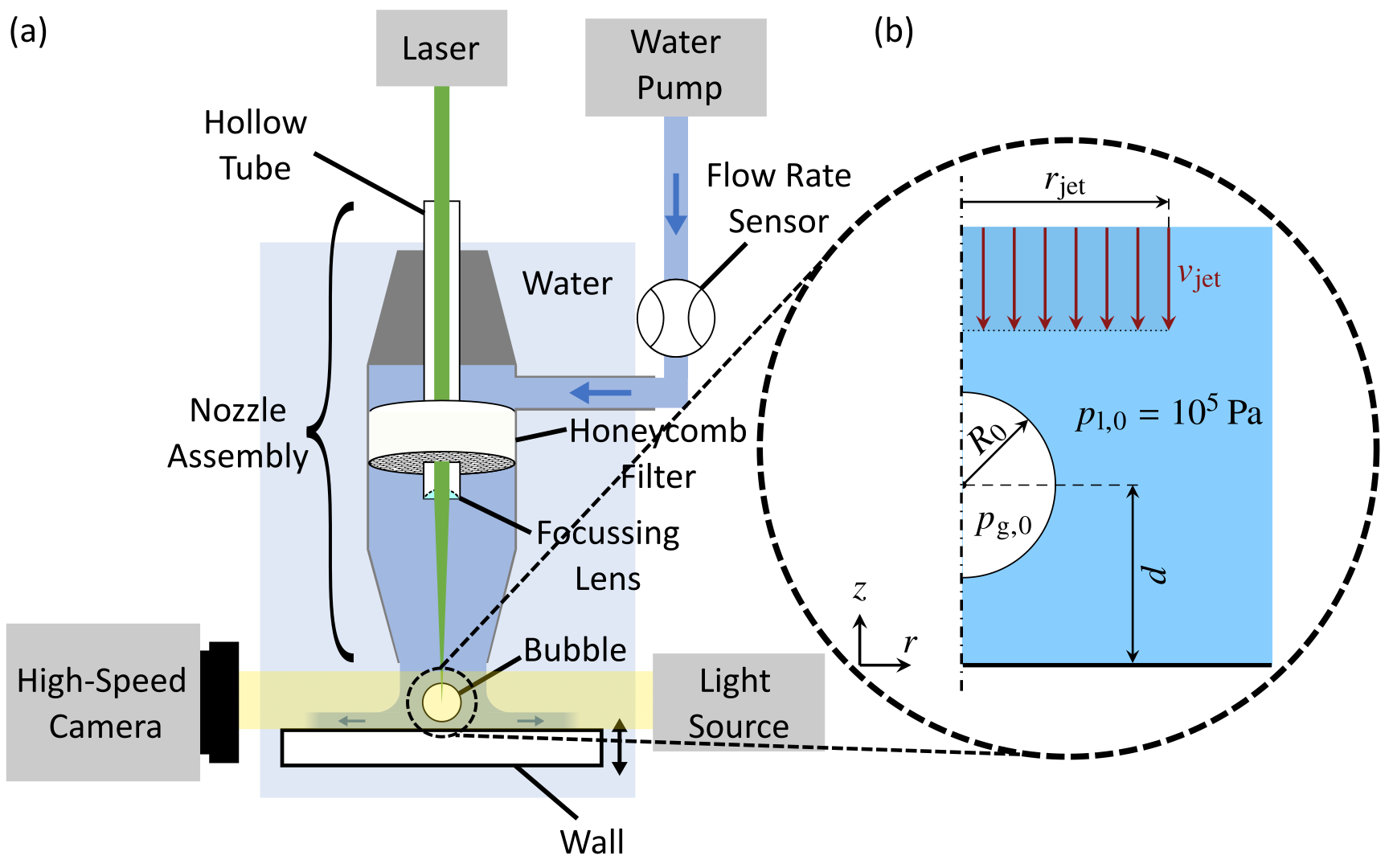}
\caption{(a) Sketch of the experimental setup. The cavitation bubble is generated by focusing the laser pulse through a lens into the wall jet. The wall jet flow is submerged to ease optical access using a high-speed camera. The liquid flow is driven with a pump and controlled with a flow rate sensor feedback loop. A honeycomb filter straightens the flow, resulting in a laminar wall jet flow. The distance of the bubble to the wall is adjusted by moving the glass plate, while the nozzle assembly remains fixed. (b) Schematic illustration of the bubble in the wall jet and the axisymmetric simulation setup (not to scale).}
    \label{fig:sketch_combined}
\end{figure}

A single, laser-induced cavitation bubble is produced in a wall jet flow at a well-controlled distance to a wall. The submerged wall jet \citep{Glauert1956} is issued from a conical nozzle and provides a laminar water stream perpendicular to the wall that spreads radially outwards, see Figure~\ref{fig:sketch_combined}, i.e., providing a stagnation point flow. The cavitation bubble is laser-seeded on the central axis of the wall jet at a variable distance from the wall. %Figure~\ref{sketch} depicts a sketch of the experimental configuration. 

%\begin{figure}
%    \centering
%    \includegraphics{simulation_setup}
%    \caption{Schematic illustration of the bubble in the wall jet and the axisymmetric simulation setup (not to scale).}
%    \label{fig:simulation_setup}
%\end{figure}

%\begin{figure}
%    \centerline{\includegraphics[width=10.5cm]{fig_exp/Sketch_Hochkant_V.png}}
%\caption{Sketch of the experimental setup. The cavitation bubble is generated by focusing the laser pulse through a lens into the wall jet. The wall jet flow is submerged to ease optical access using a high-speed camera. The liquid flow is driven with a pump and controlled with a flow rate sensor feedback loop. A honeycomb filter straightens the flow, resulting in a laminar wall jet flow. The distance of the bubble to the wall is adjusted by moving the glass plate, while the nozzle assembly remains fixed.\label{sketch}}
%\end{figure}

For convenience, the water is collected in a transparent water tank and recirculated. About 1.2 litres of water are used and pumped with a rotary vane pump (aquastream XT, Aqua Computer GmbH, Germany) through the nozzle. The nozzle is fully submerged in the water of the tank to ease observation of the bubble dynamics within the wall jet. The back part of the nozzle assembly is open to air, see Figure~\ref{fig:sketch_combined} (a). Inside the nozzle, the water passes through a honeycomb filter (stacked cylinders with $2\,$mm diameter and $30\,$mm length) to straighten the flow. The volumetric flow rate $Q$ is measured with a turbine flow meter (FCH-midi-POM, B.I.O-TECH e.K.) located between the pump and the nozzle inlet.
The average wall jet velocity is given by $v_{\mathrm{jet}}=Q/A_{\mathrm{nozzle}}$, where $A_{\mathrm{nozzle}}=\pi r_{\mathrm{nozzle}}^2$ is the cross-sectional area of the circular nozzle with inner radius of $r_{\mathrm{nozzle}}=1.5\,$mm.
In our setup, $v_{\mathrm{jet}}$ can be controlled between $0$ and $5\,$m/s using the software provided with the pump.

To generate single cavitation bubbles, one pulse from a frequency-doubled Nd:YAG laser (wavelength of $532\,$nm) is first collimated and guided through a hollow tube from the back end of the nozzle assembly. Then the laser is focused, and a bubble explosively grows as a result of the optic breakdown. Focusing is achieved using a planoconvex lens of focal length $f=10\,$mm (in air) with a diameter of $6\,$mm (Thorlabs, LA1116). The lens is bonded into the wall-facing end of a tube. The planar side of the lens is in contact with water, while the curved side is in contact with air on the hollow tube side. The distance of the lens to the nozzle exit is 11 mm. When the laser is fired, a cavitation bubble is created at a distance of $4.4\,$mm from the nozzle exit. Due to the strong deformation of the bubbles, in experiments and simulations, the volume-equivalent bubble radius was measured. The geometry of the 3D-printed nozzle (PLA plastic) went through a number of iterations to achieve a smooth water flow along the lens such that no air gets entrapped at the lens output aperture. A linear convergent nozzle together with a honeycomb flow rectifier accomplishes this requirement. In addition, there is no significant recirculation or disturbance of the stagnation flow since the bubble size is much smaller than the outer nozzle diameter (7 mm) and the solid sample. To confirm the laminar flow profile, we have used micrometre-sized particles as flow tracers together with high-speed imaging, see Appendix A.

The wall is provided by a square-shaped coverslip with side length $24\,$mm and a thickness of $0.16\,$mm. It is mounted onto a plastic frame such that a central "window" of 14 x 14 mm$^2$ remains for observation through the wall. The size of the glass sample is much larger than the nozzle diameter such that edge effects of the sample do not influence the flow in the gap between nozzle and sample surface. The distance between the laser plasma, i.e. bubble generation, and the glass wall is adjusted with a translation stage. 
%The nozzle is fixed in the water tank, thus once the laser is aligned to pass centrally through the hollow tube, the distance $d$ between the laser focus and the boundary is adjusted using the translation stage. 
The bubble dynamics is recorded with a high-speed camera (Shimadzu HPV-X2) imaging through the water-filled tank and the discharged wall jet flow from the nozzle. As a light source, we used an intense, pulsed LED (LED-P40, SMETec).

\subsection{Simulation methods}
Numerical simulations complement the experimental recordings.
The primary aim of these simulations is to investigate the differences in wall shear stress and pressure generated by a collapsing bubble in a wall jet flow, as compared to in an initially stagnant body of liquid. Experimentally, these quantities are inaccessible due to the required temporal and spatial resolutions.

The considered compressible two-phase flow is governed by the continuity equation
\begin{equation}
    \frac{\partial \rho}{\partial t} + \boldsymbol{\nabla} \cdot (\rho \mathbf{u}) = 0
\end{equation}
and the momentum equation
\begin{equation}
    \frac{\partial \rho \mathbf{u}}{\partial t} + \boldsymbol{\nabla} \cdot (\rho \mathbf{u} \otimes \mathbf{u}) = -\boldsymbol{\nabla} p + \boldsymbol{\nabla} \cdot \left[ \mu \left( \boldsymbol{\nabla} \mathbf{u} + \boldsymbol{\nabla} \mathbf{u}^\mathrm{T} \right) - \frac{2}{3} \mu \left(\boldsymbol{\nabla} \cdot \mathbf{u} \right) \mathbf{I} \right] ,
\end{equation}
where $t$ represents time, $\rho$ is the mass density of the fluid, $\mu$ is the dynamic viscosity of the fluid, $\mathbf{u}$ is the velocity of the flow, $p$ is the pressure and $\mathbf{I}$ is the identity tensor. 
An algebraic Volume-of-Fluid (VOF) method \citep{Denner2018b} is adopted to model the gas-liquid interface and distinguish the gas and the liquid.
The governing conservation laws are closed using the Noble-Abel stiffened gas (NASG) equation of state \citep{LeMetayer2016} in polytropic form, with the density and speed of sound defined as
\begin{eqnarray}
    \rho &=& \frac{K (p+\Pi)^{1/\kappa}}{1+ b K (p + \Pi)^{1/\kappa}},\\
    c &=& \sqrt{\kappa \, \frac{p+\Pi}{\rho (1-b \rho)}},
\end{eqnarray}
respectively,
where $\kappa$ is the polytropic exponent, $\Pi$ is a pressure constant and $K=\rho_0 / [(p_0 + \Pi)^{1/\kappa} (1-b\rho_0)]$ defines a constant reference state based on a reference pressure $p_0$ and density $\rho_0$. 

We are choosing the polytropic form of the NASG equation of state because we are not focussing on the details of the decay of the emitted pressure pulses that may (or may not) form shock fronts as they propagate away from the bubble, and since the Peclet number associated with the liquid jet that pierces the bubble during collapse is high, $\mathrm{Re} > 10^3$, we do not consider heat transfer to be an important factor for our simulations. Using such a polytropic fluid model has been demonstrated to be suitable even for the prediction of complex bubble behaviour, such as wall-bounded cavitation \citep{Koch2016,Zeng2018}. Following previous work we are neglecting surface tension and justify this simplification by (i) the relatively large bubble size with a volume-equivalent diameter $> 1 \, \mathrm{mm}$ at maximum expansion, and (ii) the high Weber number, $\mathrm{We} > 10^3$, of the liquid jet that pierces the bubble during collapse.
 
The liquid is assumed to be water, with $\kappa_l = 1.186 $, $\Pi_l = 7.028 \times 10^8 \, \mathrm{Pa}$, $b_l = 6.61 \times 10^{-4} \, \mathrm{m}^3/\mathrm{kg}$, $\rho_{0,l} = 957.74 \, \mathrm{kg/m}^3$, $c_{0,l} = 1540.2 \, \mathrm{m/s}$ and $p_{0,l} = 10^5 \, \mathrm{Pa}$, as previously proposed by \cite{LeMetayer2016}, and $\mu_l = 10^{-3} \, \mathrm{Pa \, s}$. The bubble content is taken to be noncondensable air, modelled as an ideal gas ($\Pi_g = 0$, $b_g=0$), with $\kappa_g = 1.4$, $\rho_{0,g} = 1.2 \, \mathrm{kg/m}^3$, $p_{0,g} = 10^5 \, \mathrm{Pa}$ and $\mu_g = 1.82 \times 10^{-5} \, \mathrm{Pa \, s}$.
The governing equations are discretised using a second-order finite-volume method and solved using a fully-coupled implicit pressure-based algorithm \citep{Denner2018b,Denner2020}.

The computational setup is illustrated schematically in Figure \ref{fig:sketch_combined} (b).
The axisymmetric simulations are carried out in a $0.25 \, \mathrm{m} \times 0.25 \, \mathrm{m}$ computational domain, which is sufficiently large such that the boundary conditions do not influence the bubble dynamics or the flow field in the vicinity of the bubble in the considered time frame. 
The bubble is initialised at a distance $d$ from the wall with initial gas pressure $p_\mathrm{g,0}=54.5 \, \mathrm{MPa}$ and initial radius $R_0=84 \, \mu\mathrm{m}$, tuned to achieve a desired maximum radius $R_\mathrm{max}$. Based on the experimental setup, the wall jet flow is represented by a uniform flow with velocity magnitude $v_\mathrm{jet}$ and radius $r_\mathrm{jet} = 1500 \, \mu \mathrm{m}$. 

\begin{figure}
    \centering
    \includegraphics[width=\linewidth]{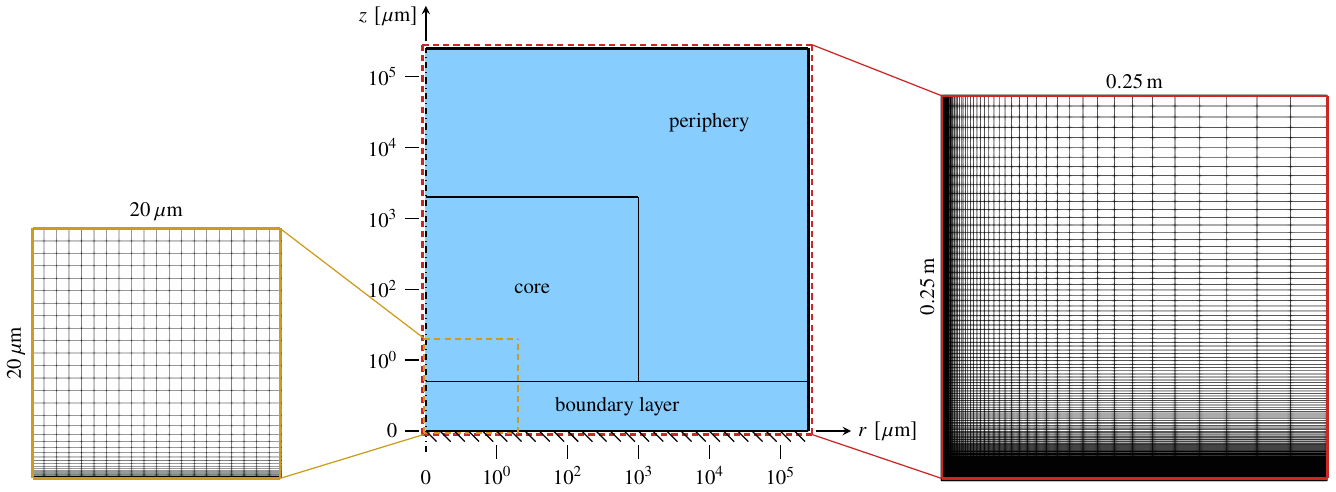}
    \caption{Schematic illustration of the computational domain with the 3 primary mesh regions (sketched on a logarithmic scale), with a close up of the computational mesh near the wall on the left and the complete computational mesh on the right.}
    \label{fig:mesh_setup}
\end{figure}

The mesh resolution required to resolve the considered cases adequately is primarily governed by the diameter of the liquid jet penetrating the bubble, $\gtrsim 10 \, \mu \mathrm{m}$, and the velocity gradient normal to the wall, $\mathcal{O}(10^7)-\mathcal{O}(10^9)  \, \mathrm{s}^{-1}$, associated with the generated wall shear stresses. The computational mesh is static (i.e.~does not adapt) and separated into 3 primary regions, as illustrated in Figure \ref{fig:mesh_setup}: the boundary layer close to the rigid wall, the core region with the bubble and stagnation flow, and the periphery which mainly exists to avoid pressure waves reaching the domain boundaries. Note that, due to the large range of lengthscales, the computational domain is sketched on a logarithmic scale in Figure \ref{fig:mesh_setup}.
In line with previous studies on wall-bounded bubble collapse \citep{Zeng2018, Lechner2020, Gonzalez-Avila2021, Mifsud2021}, the core region of the domain ($r \leq 1000 \, \mu \mathrm{m}$ and $z \leq 2000 \, \mu \mathrm{m}$) is resolved with a mesh spacing of $\Delta x_0 = 1 \, \mu \mathrm{m}$ and the mesh near the wall ($z < 5  \, \mu \mathrm{m}$) is gradually refined, such that the centres of the layer of cells closest to the wall are located at a distance of only $12.5 \, \mathrm{nm}$ from the wall. 
% As discussed in more detail in Section \ref{sec:results_sim}, such a high resolution is required to resolve the steep velocity gradients observed in the considered cases adequately. 
In the periphery region of the domain ($r > 1000 \, \mu \mathrm{m}$ and $z > 2000 \, \mu \mathrm{m}$), the mesh is gradually coarsened. 
Similar to our previous work \citep{Gonzalez-Avila2021}, the adaptively chosen time-step is $\Delta t = \mathrm{Co} \, \Delta x / |\mathbf{u}|$, where $\mathrm{Co} = 0.7$ is the applied convective Courant number.

The employed numerical methods and case setup were validated successfully against experiments of laser-induced bubble expansion and collapse near walls in our previous work, regarding the bubble evolution and shape as well as the acoustic emissions of the bubble collapse \citep{Gonzalez-Avila2021}, and regarding the minimum thickness of the thin liquid film remaining between bubble and wall \citep{Reuter2019high,Denner2020a}. Furthermore, the numerical methods have been compared favourably to experimental measurements and numerical predictions of shock-driven bubble collapse \citep{Denner2018b} and to the Gilmore model for a Rayleigh collapse \citep{Denner2023}. The specific resolution requirements of the simulated bubble dynamics are discussed alongside the simulation results in Section \ref{sec:results} and representative results obtained with different mesh resolutions are reviewed in Section \ref{sec:summary and discussion}.

\subsection{Scaling with bubble size}
Here, we consider bubbles with maximum radii ranging between $R_\mathrm{max}=480\,\mu$m and $716\,\mu$m. To ease the discussion, we show that for a bubble in the considered wall jet, the same scaling as for a Rayleigh bubble holds.
%Before presenting the results in Section \ref{sec:results}, we discuss some preliminary considerations, concerning the influence of the wall jet and its dependency on the bubble size. 

The potential energy of a laser-induced cavitation bubble with negligible vapour pressure ($p_\mathrm{vap} \ll p_\infty$) is customarily approximated as \citep{Liang2022}
\begin{equation}
    E_\mathrm{pot} = \frac{4}{3} \pi R_\mathrm{max}^3 p_\infty  \propto R_\mathrm{max}^3 p_\infty,
\end{equation}
%where $p_\mathrm{vap}$ is the vapour pressure of the gas. 
We estimate the kinetic energy imparted on the bubble by the wall jet as the product of the dynamic pressure of the wall jet and a suitably defined reference volume,
\begin{equation}
    E_\mathrm{jet} = \frac{\rho_{0,l}}{2} v_\mathrm{jet}^2 V_\mathrm{ref}.
\end{equation}
We further assume that the wall jet acts on the upper hemisphere of the bubble, with a projected area of $A = \pi R^2_\mathrm{max}$. Since the distance over which the bubble is advected by the wall jet during the bubble lifetime is negligible ($R_{max} \gg v_{jet} T_C$), $R_\mathrm{max}$ is the reference length scale of the problem. Thus the reference volume is then defined as
%With $R_\mathrm{max}$ as the reference length scale of the problem, the reference volume is then defined as
\begin{equation}
V_\mathrm{ref} = A \, R_\mathrm{max}, \label{eq:Vjet}    
\end{equation}
such that $E_\mathrm{jet} \propto \rho_{0,l} v_\mathrm{jet}^2 R_\mathrm{max}^3$.
For simplicity, we do not take surface tension and viscosity into account, since a collapsing cavitation bubble is an inertia-driven process, see for example \citet{reuter2022rayleigh}.
% \begin{equation}
%     V_\mathrm{jet} = \pi R_\mathrm{max}^2 v_\mathrm{jet} T_\mathrm{C},
% \end{equation}
% and $T_\mathrm{C}$ is the collapse time of the bubble. When assuming a Rayleigh collapse, expansion and collapse phases are symmetrical, and we can define the bubble lifetime as $T_\mathrm{L}=2T_\mathrm{C}$. 
Taking the ratio of $E_\mathrm{jet}$ and $E_\mathrm{pot}$ yields
\begin{equation}
\frac{E_\mathrm{jet}}{E_\mathrm{pot}} \propto \frac{\rho_{0,l} v_\mathrm{jet}^2}{p_\infty}.
\end{equation}
Consequently, neglecting surface tension and viscous dissipation, the influence of the wall jet can be assumed to be independent of the bubble size. Therefore, the presented results are, in first approximation, independent of the bubble size.

This allows us to normalise the times on the bubble lifetime $T_\mathrm{L}$, i.e. the time between bubble nucleation and first collapse (minimum volume) as $T_{\mathrm{L}}\propto R_{\mathrm{max}}$, to facilitate the comparison of similarly sized bubbles.
%To facilitate the comparison of similarly sized bubbles, times are given normalised on the bubble lifetime $T_\mathrm{L}$, i.e. the time between bubble nucleation and first collapse (minimum volume) as $T_{\mathrm{L}}\propto R_{\mathrm{max}}$.  

\section{Results}
\label{sec:results}

The results are organised by reporting the effect of the flow velocity of the wall jet, $v_\mathrm{jet}$, on the bubble dynamics for three different stand-off distances $\gamma$. The first stand-off is (1) far from the wall at a large distance of $\gamma \approx 1.7$. Here, the bubble shows a pinch-off, as shown in previous studies for cavitation bubbles in a stagnation flow \citep{Starrett1982,blake_taib_doherty_1986,robinsonblake1994,Blake2015}. The second stand-off is (2) at an intermediate distance of $\gamma \approx 0.7$, which shows a needle-like high-speed jet, as observed in stagnant liquid for smaller stand-off distances, $\gamma \lessapprox 0.2$ \citep{Lechner2019,Reuter2021,Bussmann2023}. Last, a cavitation bubble (3) representing the small stand-off distance dynamics ($\gamma \approx 0.4$) is presented. Here, the bubble pinches off directly above the wall, resulting in a lift-off of the bubble before it collapses. This is probably the most interesting case regarding the bubble-wall interaction, since a potentially harmful collapse in direct contact with the wall can be avoided. 
For each stand-off distance, high-speed imaging and numerical results of the bubble dynamics are shown and analysed, and the spatio-temporal wall shear stress and wall pressure distributions are discussed. Note that the given values may be considered as lower bounds of these quantities, which is discussed in more detail in Section \ref{sec:summary and discussion}.

\subsection{Large stand-off distance}

\begin{figure}
\centerline{\includegraphics[width=13cm]{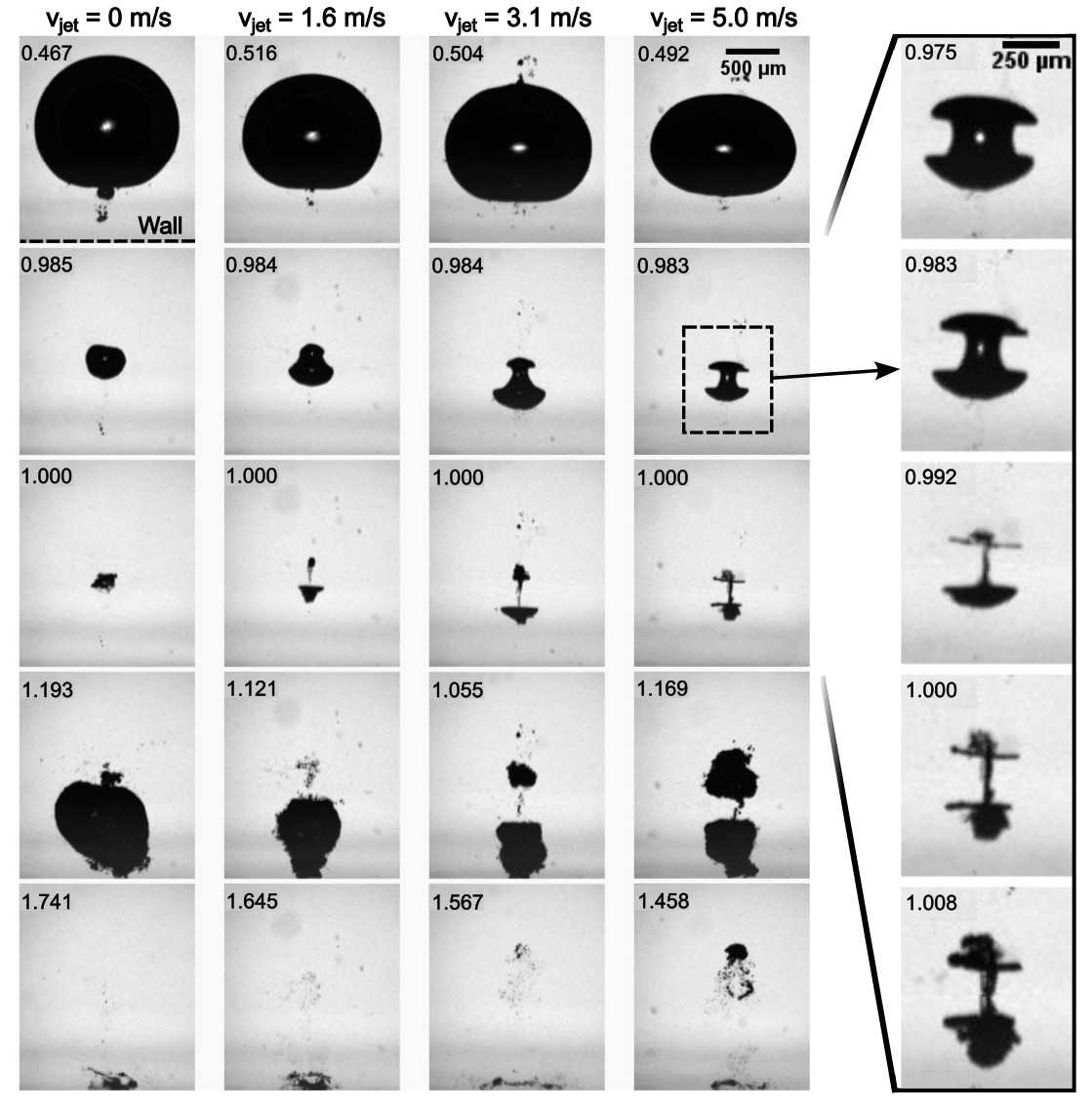}}
\caption{High-speed imaging of bubble dynamics for four different wall jet velocities $v_{jet}$ at large stand-off distances ($\gamma=1.73,1.78,1.56,1.88$, from left to right). For sufficiently large wall jet velocities, the bubble pinches off and two axial jets develop in opposite directions. Times indicated in each tile are normalised to the bubble lifetime. The respective lifetimes are $T_\mathrm{L}=135,124,127,118\,\mu$s. The bottom of each frame coincides with the wall, as sketched in the first tile. The rightmost column shows the bubble splitting and collapse for the last case in more detail. A video of the dynamics at $v_{jet}=5\,$m/s can be found in the Supplementary Material as Movie 1.}
\label{fig: v_Dependence large gamma}
\end{figure}

\begin{figure}
    \centerline{
    \includegraphics[width=14cm]{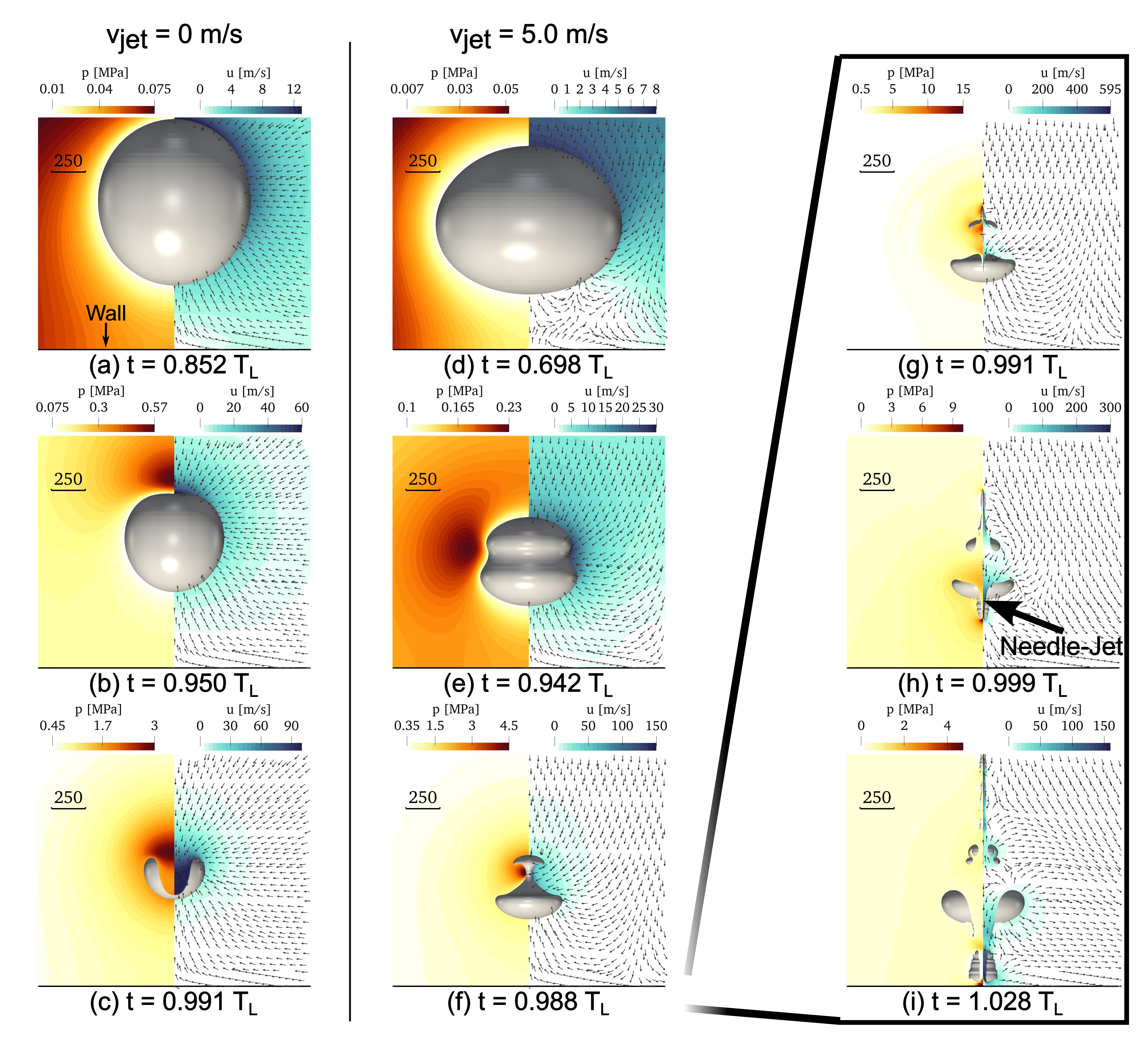}}
    \caption{Bubble shape, pressure (left half) and velocity field (right half) at different time instances during the collapse of the bubble initially located at $d = 1200\, \mu \mathrm{m}$ ($\gamma = 1.53$ and $1.75$, respectively). The arrows in the right half of each figure indicate the direction of the flow.
    (a) to (c) In quiescent water ($v_\mathrm{jet} = 0 \, \mathrm{m/s}$). The lifetime of this bubble is $T_\mathrm{L} = 164.4 \, \mu \mathrm{s}$.
    (d) to (i) Subject to a wall jet with $v_\mathrm{jet} = 5 \, \mathrm{m/s}$. The lifetime of this bubble is $T_\mathrm{L} = 143.4 \, \mu \mathrm{s}$. The scale bar corresponds to $250 \, \mu \mathrm{m}$.}
    \label{fig:sim_h1200_p-u}
\end{figure}

\begin{figure}
\centering
    \subfloat[$v_\mathrm{jet} = 0 \, \mathrm{m/s}$, $\gamma = 1.53$]{\includegraphics[width=0.85\linewidth]{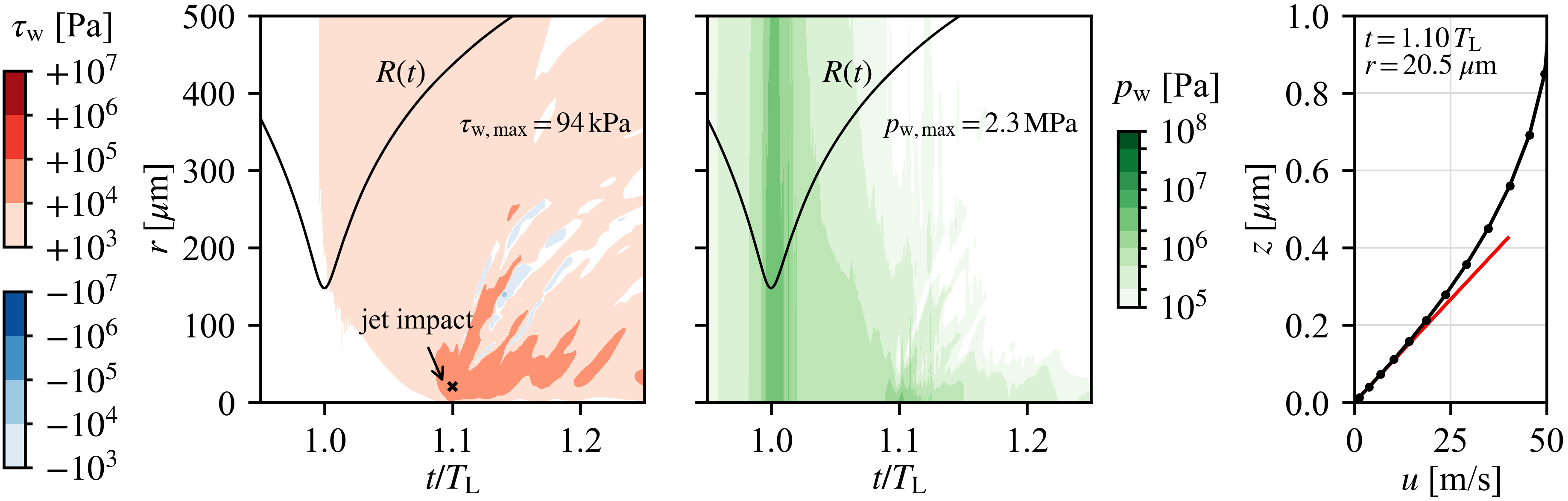}} \\
    \subfloat[$v_\mathrm{jet} = 5 \, \mathrm{m/s}$, $\gamma = 1.75$]{\includegraphics[width=0.85\linewidth]{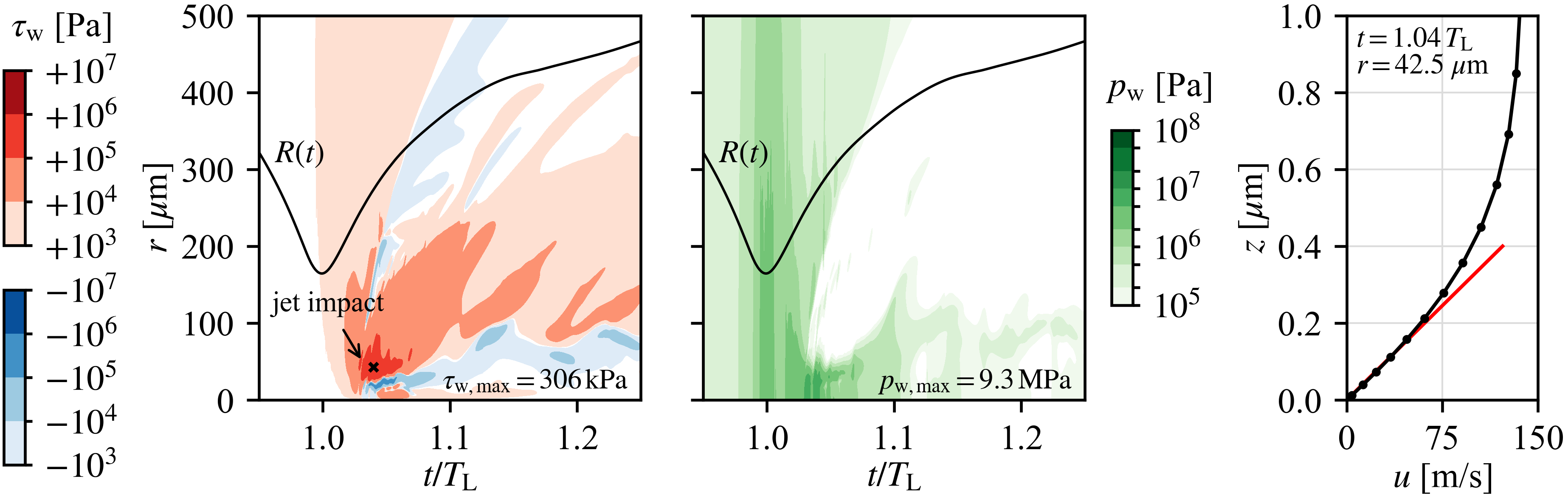}}
    \caption{Space-time plots of the wall shear stress $\tau_\mathrm{w}$ and the wall pressure $p_\mathrm{w}$, and profile of the radial velocity $u$ of the liquid at the location of the highest wall shear rate, of the bubble initially located at $d = 1200 \, \mu \mathrm{m}$. In the space-time plots, the black line shows the volume-equivalent bubble radius $R(t)$ and, with respect to $\tau_\mathrm{w}$,  red (blue) areas indicate a radially outward (inward) going flow. The location of maximum wall shear stress is indicated with a bold cross in the space-time plots. In the plots of the velocity profiles, the red line represents the velocity gradient associated with $\tau_\mathrm{w,max}$, and the black dots show the locations of the cell centers of the applied computational mesh.}
    \label{fig:bl_h1200}
\end{figure}

Figure \ref{fig: v_Dependence large gamma} depicts the typical bubble dynamics for large stand-off distances. In the selected series, $\gamma$ is between $1.56$ and $1.88$ and the glass wall is located at the bottom of each frame. Each column shows five snapshots from one experiment, showing the bubble at maximum expansion, the bubble just before and just after the collapse (from top to bottom), as well as the instance when the bubble makes contact with the wall and the second collapse. Each column in Figure \ref{fig: v_Dependence large gamma} shows one experiment with a specific wall jet velocity $v_{\mathrm{jet}}$. Four wall jet velocities are shown, increasing from left to right. 

The first column without jet flow ($v_\mathrm{jet}=0\,$m/s) depicts the well-known jetting behaviour of a bubble nucleated at a large stand-off distance (here $\gamma = 1.73$) from the wall, as previously reported, for instance, by \citet{Philipp1998}. At maximum expansion, $t=0.467\,T_\mathrm{L}$, the bubble is rather spherical but with some small bubbles at the wall-near pole. Here, the usage of a small lens with a rather small opening angle to focus the laser results in some aberration and multiple breakdowns. 
During collapse, the presence of the wall impedes the radial inflow from below, and a pressure gradient forms, with a higher pressure at the top. As a result, the bubble's centroid moves towards the wall \citep{Ellis1966} and the bubble develops a jet that passes through the bubble interior toward the wall. The specific illumination applied in the experiments does not reveal this jet flow, but it is visible by the small axial indentation of the upper pole just before the collapse. At time $t=1.193\,T_\mathrm{L}$ the main bubble reaches the wall while re-expanding. Later, it collapses a second time, now in proximity to the wall at $t=1.741\,T_\mathrm{L}$. 

The simulated dynamics of the corresponding bubble is shown in Figure~\ref{fig:sim_h1200_p-u}(a)-(c). Here the jet formation is clear and jet velocities reach up to $100 \, \mathrm{m/s}$, which is in close agreement with previous studies \citep{Zeng2018, Gonzalez-Avila2021}. 
Stresses on the wall are given in Figure \ref{fig:bl_h1200}(a). 
The peak wall pressure is reached at $t = 1.003 \, T_\mathrm{L}$ and associated with the bubble collapse, rather than the jet impact pressure at $t = 1.101 \, T_\mathrm{L}$, in agreement with our previous work \citep{Gonzalez-Avila2021}. However, as the jet impacts the wall, it generates significant wall shear stresses, with a maximum of $94 \, \mathrm{kPa}$.

Let us now look at how the wall jet flow modifies the bubble dynamics, considering the case of the largest wall jet flow velocity $v_{\mathrm{jet}}=5\,$m/s first. In comparison to the stagnant case, the bubble takes an ellipsoidal shape with the major axis aligned parallel to the wall, i.e. it is flattened at the bottom, see the first row. A video of the corresponding bubble is available as Movie 1 of the Supplementary Material. As the bubble collapses, instead of an indentation in the axial direction from the regular axial jet flow, a planar convergent jet indents the bubble from the sides and results in the horizontal kink in the second row. As this flow converges towards the axis of symmetry, two oppositely directed jets are ejected axially, see third row. More insights into this process are gained from the flow and pressure fields of Figure~\ref{fig:sim_h1200_p-u}(d)-(f) and the snapshot series at smaller time intervals showing the jet formation and piercing process in Figure~\ref{fig:sim_h1200_p-u}(g)-(i) as well as the rightmost column of Figure~\ref{fig: v_Dependence large gamma}. 
The convergent planar flow 
% results 
% in a peak stagnation pressures that spike up to $\approx50 \, \mathrm{MPa}$
%The convergent planar flow results in a peak stagnation pressure of around $50 \, \mathrm{MPa}$  
% at $t = 0.988 \, T_\mathrm{L}$ in Figure~\ref{fig:sim_h1200_p-u}(f) 
% and 
results in an axial needle-like jet with a radius of the order of $6 \, \mu \mathrm{m}$, indicated by an arrow in Figure~\ref{fig:sim_h1200_p-u}(h). 
The jet directed toward the wall reaches a maximum velocity of $\approx 300 \, \mathrm{m/s}$, a threefold increase compared to the liquid jet generated in the corresponding case with no wall jet flow ($v_\mathrm{jet}=0\,$m/s). Upon impact on the wall, the jet generates a maximum wall shear stress of $306 \, \mathrm{kPa}$, see Figure~\ref{fig:bl_h1200}(b), which is roughly three times higher than in the stationary case. Due to the formation of a stagnation point around the region where the axis of symmetry intersects the wall, the shear stresses are smaller there \cite{reuter2018electrochemical}. Also, some negative shear rates are seen from splashing and recirculations of the complex flow after the jet impacts on the wall. The maximum pressures at the wall reach about $p_{\mathrm{max}}=9.3 \, \mathrm{MPa}$ (a fourfold increase compared to $v_\mathrm{jet}=0\,$m/s) and coincide with the jet impact. They are not generated from the shockwaves emitted during the collapse. The impact velocity of the jet can be approximated as $u=135 \, \mathrm{m/s}$ as the maximum pressure on the wall corresponds to the stagnation pressure of the jet, $p_{\mathrm{max}}=0.5~\rho_{0,l}~u^2$ \citep{Gonzalez-Avila2021}.

%From the maximum pressure on the wall, the impact velocity of the jet can be approximated as $u=135 \, \mathrm{m/s}$ using the incompressible water hammer pressure approximation: $p_{\mathrm{max}}=\rho c u$.

The pressure distribution in Figure~\ref{fig:sim_h1200_p-u}(e) reveals a larger pressure at the wall distal side around the bubble, which explains why, after pinch-off, the upper bubble collapses before the lower bubble. 
%The pressure distribution in Figure~\ref{fig:sim_h1200_p-u}(e) the wall distal side around the bubble, which explains why, after pinch-off, the upper bubble collapses before the lower bubble. 
For increasing wall jet velocity, the bubble becomes more ellipsoidal, see Figure~\ref{fig: v_Dependence large gamma} first row. Furthermore, while at $v_\mathrm{jet}=5\,$m/s the pinch-off occurs about equatorial, which results in the bubble splitting into two about equally sized volumes, for lower wall jet velocities the pinch-off occurs closer to the upper bubble pole, which, in turn, results in an uneven splitting of the bubble. A smaller bubble is produced above and a larger bubble below the pinch-off region, as observed comparing the second and fourth row in Figure~\ref{fig: v_Dependence large gamma}.

\subsection{Intermediate stand-off distance}

\begin{figure}
    \centerline{\includegraphics[width=13cm]{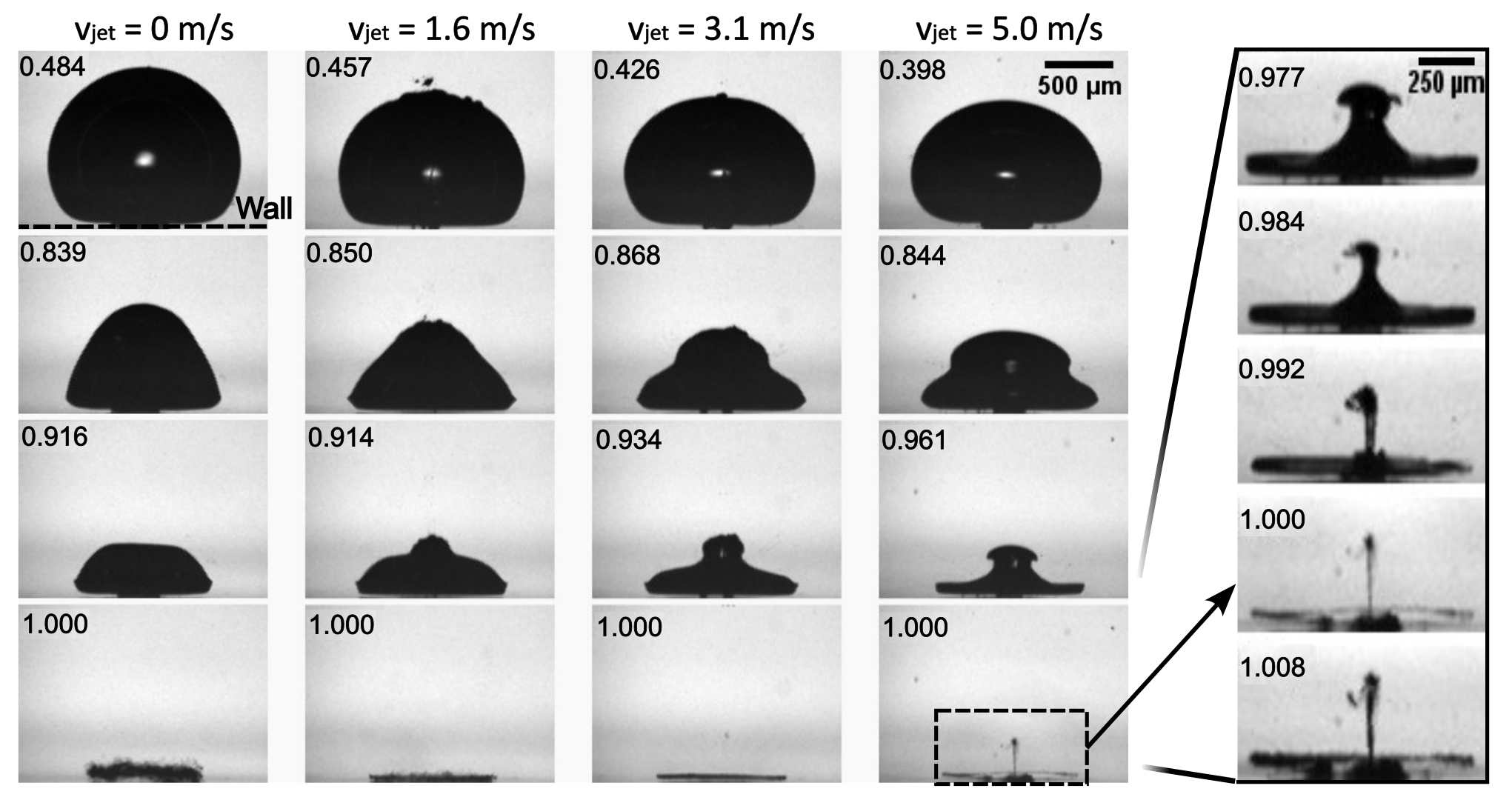}}
\caption{Bubble dynamics with increasing wall jet velocity at intermediate stand-offs ($\gamma=0.74,0.72,0.78,0.78$, from left to right). The wall extends along the bottom of the frames. 
For the fastest wall jet flow, a high-speed needle-like jet occurs.
%The wall jet velocity $v_\mathrm{jet}$ is indicated above each column. 
Times indicated in each tile are normalised to the bubble lifetime. The respective lifetimes are $T_\mathrm{L}=155, 140, 136, 128  \, \mu \mathrm{s}$ (from left to right).
 The collapse for the case $v_\mathrm{jet}=\mathrm{max}$ is studied in further detail in the rightmost column. The corresponding video of the bubble at $v_{jet}=5\,$m/s can be found in the Supplementary Material as Movie 2.
\label{fig: v_Dependence intermediate gamma}}
\end{figure}

\begin{figure}
    \centerline{\includegraphics[width=14cm]{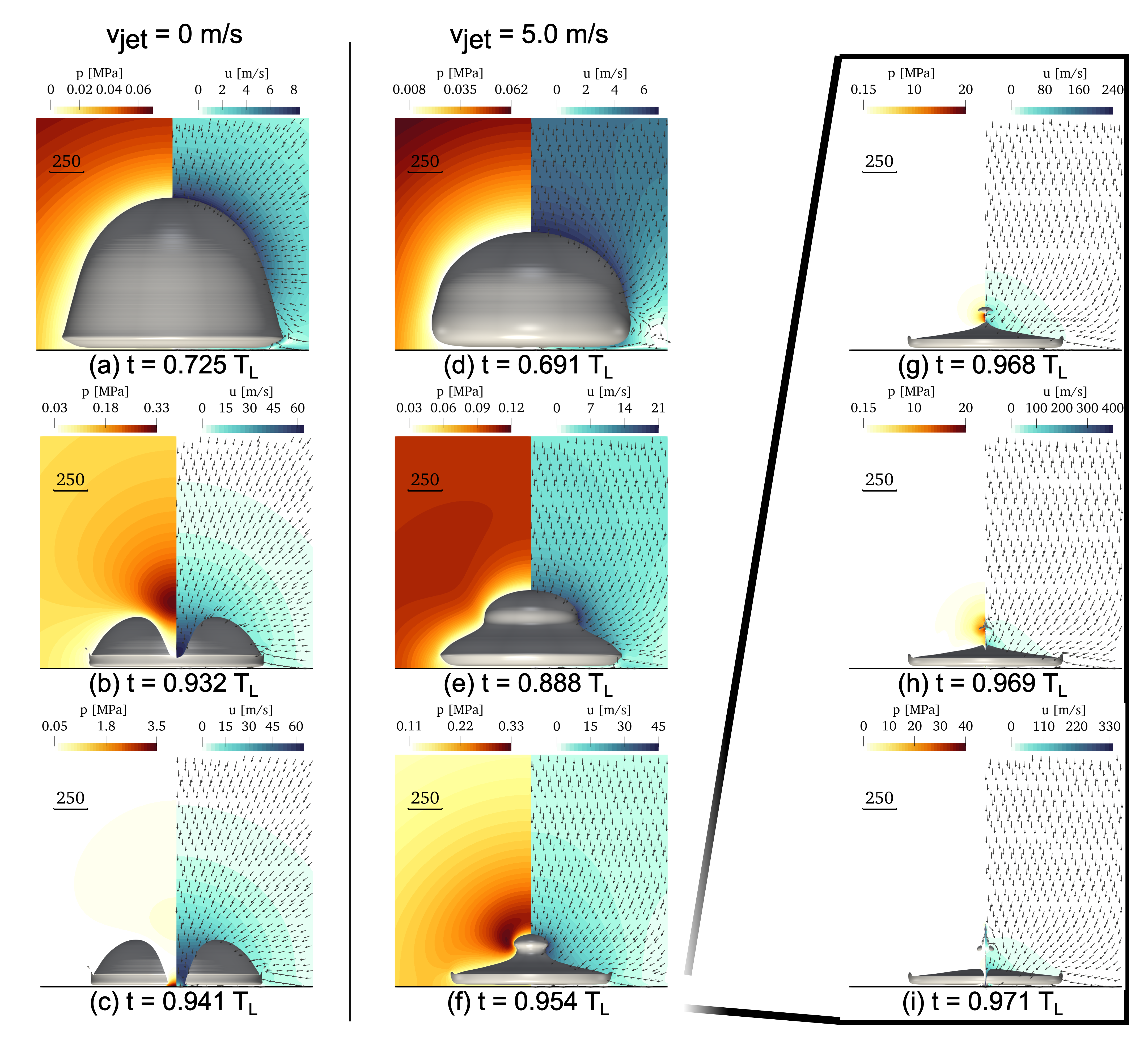}}
    \caption{Bubble shape, pressure and velocity field at different time instances during the collapse of the bubble initially located at $d = 490 \, \mu \mathrm{m}$ ($\gamma = 0.63$ and $0.73$, respectively).
    (a) to (c) In quiescent water ($v_\mathrm{jet} = 0 \, \mathrm{m/s}$). The lifetime of this bubble is $T_\mathrm{L} = 179.3 \, \mu \mathrm{s}$.
    (d) to (i) Subject to a wall jet with $v_\mathrm{jet} = 5 \, \mathrm{m/s}$. The lifetime of this bubble is $T_\mathrm{L} = 152.0 \, \mu \mathrm{s}$. The scale bar corresponds to $250 \, \mu \mathrm{m}$.}
    \label{fig:sim_h0490_p-u}
\end{figure}

\begin{figure}
\centering
    \subfloat[$v_\mathrm{jet} = 0 \, \mathrm{m/s}$, $\gamma = 0.63$]{\includegraphics[width=0.85\linewidth]{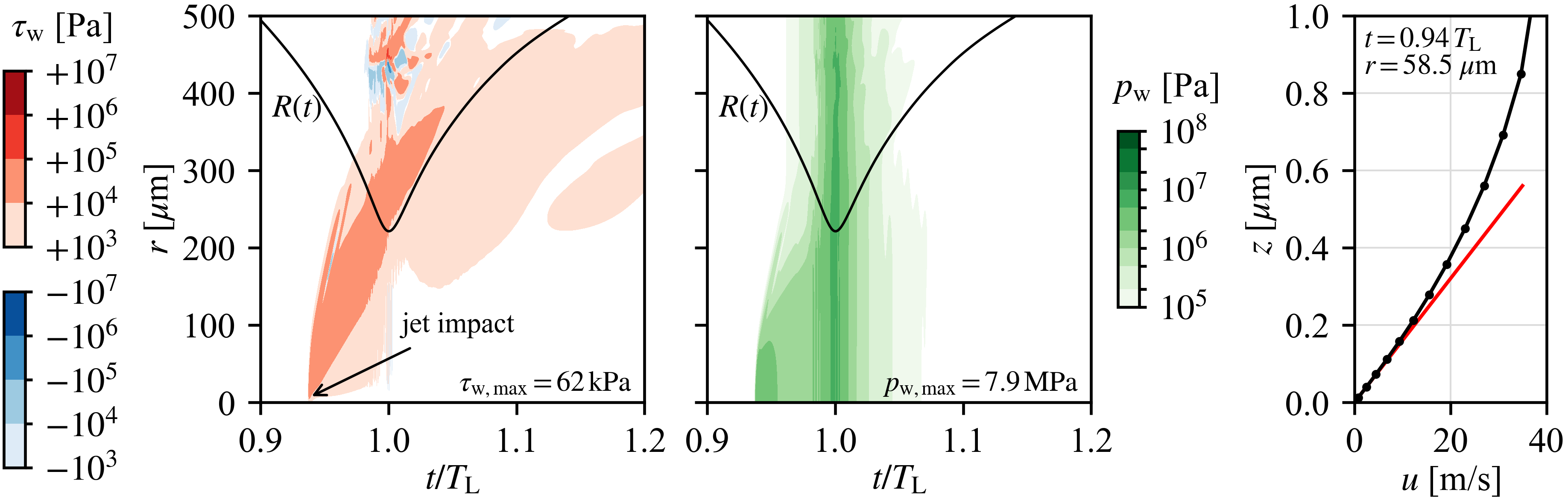}} \\
    \subfloat[$v_\mathrm{jet} = 5 \, \mathrm{m/s}$, $\gamma = 0.73$]{\includegraphics[width=0.85\linewidth]{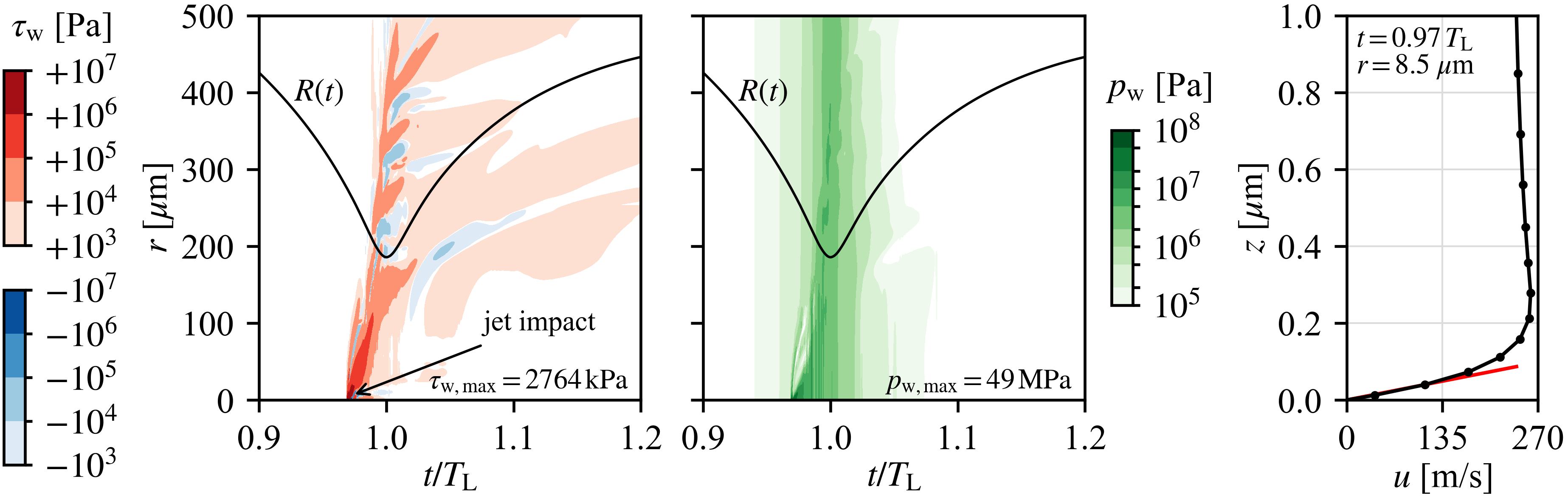}}
    \caption{Space-time plots of the wall shear stress $\tau_\mathrm{w}$ and the wall pressure $p_\mathrm{w}$, and profile of the radial velocity $u$ of the liquid at the location of the highest wall shear rate, of the bubble initially located at $d = 490 \, \mu \mathrm{m}$. In the space-time plots, the black line shows the volume-equivalent bubble radius $R(t)$ and, with respect to $\tau_\mathrm{w}$, red (blue) areas indicate a radially outward (inward) going flow. In the plots of the velocity profiles, the red line represents the velocity gradient associated with $\tau_\mathrm{w,max}$, and the black dots show the locations of the cell centres of the applied computational mesh.}
    \label{fig:bl_h0490}
\end{figure}

Figure \ref{fig: v_Dependence intermediate gamma} shows the evolution of the bubble dynamics for increasing wall jet flow velocities at intermediate stand-off distances of $\gamma\approx 0.75$.
%, with $\gamma$ ranging from 0.72 to 0.78. 
For each velocity the bubble is shown in four phases: 1) the moment of maximum expansion, 2) in the collapse phase when being increasingly deformed by the inflow, 3) the shape before, and 4) after the bubble collapse.

Again, in the first column, a collapsing bubble in quiescent liquid is shown. At $t=0.484\,T_\mathrm{L}$, the bubble has reached approximately its maximum expansion. At $t=0.916\,T_\mathrm{L}$ the bubble is pierced by a rather wide jet, which is easily visible in the simulation result shown in Figure~\ref{fig:sim_h0490_p-u}(b). The velocity of this jet is $\approx 65 \, \mathrm{m/s}$ and results in a toroidal collapse of the bubble in Figure~\ref{fig: v_Dependence intermediate gamma} at $t = 1.000 \, T_\mathrm{L}$. Upon impact at $t \approx 0.938 \, T_\mathrm{L}$ on the wall, the jet generates a wall shear stress of up to $62 \, \mathrm{kPa}$, see Figure~\ref{fig:bl_h0490}(a). The torus collapse close to the wall generates even larger shear stresses of up to $157 \, \mathrm{kPa}$ at $r \approx 450 \, \mu \mathrm{m}$, further away from the location of the jet impact. Again, recirculations 
% and splashing 
result in negative shear stresses.
The shockwaves emitted at the collapse result in the vertically extended green region at $t = 1.000 \, T_\mathrm{L}$ in Figure~\ref{fig:bl_h0490}(a). 

When adding a wall jet flow of $v_{\mathrm{jet}}=5~\mathrm{m/s}$ (fourth column), similarly to the large $\gamma$ case, the bubble assumes an ellipsoidal shape. The maximum extension in axial direction is already reached at $t=0.398\,T_\mathrm{L}$ and the bubble is visibly pushed downward by the wall jet. As a result, the wall-near bottom of the bubble still extend along the wall, while the more distal parts already collapse (see Movie 2 of the Supplementary Material). Again, a planar inflow shapes a kink into the bubble, which can be observed clearly in the simulations in Figure~\ref{fig:sim_h0490_p-u}(d)-(f). 
This bubble shape is very similar to a bubble in the stagnant liquid case without a wall jet flow but at a much smaller stand-off when forming a needle-like high-speed jet \citep{Lechner2019,Reuter2021,Bussmann2023}. Figure~\ref{fig:sim_h0490_p-u}(d)-(i) confirms the formation of a thin needle-like jet, with peak velocities of $300-400 \, \mathrm{m/s}$. This needle jet yields a maximum wall shear stress of $2764 \, \mathrm{kPa}$ as shown in Figure~\ref{fig:bl_h0490} (b) at $t\approx 0.97 \, T_\mathrm{L}$, which is $44$ times higher compared to the conventional jetting case ($v_{\mathrm{jet}}=0 \, \mathrm{m/s}$). Despite this very large wall shear stress, the associated velocity gradients are resolved adequately by the employed computational mesh, as evident by the velocity profile that coincides with the maximum wall shear stress on the right-hand side of Figure~\ref{fig:bl_h0490}(b). The maximum wall pressure of $49 \, \mathrm{MPa}$ (a sixfold increase compared to the stationary case) is generated by the impact of this needle jet, see Figure~\ref{fig:bl_h0490}(b), from which the impact velocity of the needle jet on the wall can be estimated as $\approx 310 \, \mathrm{m/s}$. 
% In contrast to the quiescent case the jet impact peak pressures are larger than the collapse shockwave pressures (****), though the high jet impact pressure are only reached within a radius of $\approx 20 \, \mu$m.
In contrast to the large $\gamma$ case, now only an insignificant flow in upward direction is generated.

From the evolution of the bubble shape with decreasing wall jet flow velocities in Figure~\ref{fig: v_Dependence intermediate gamma}, it can be seen how the bubble becomes less ellipsoidal and the kink less pronounced. 
At $v_{\mathrm{jet}}=3.1\, \mathrm{m/s}$ and $v_{\mathrm{jet}}=5.0\, \mathrm{m/s}$ in the last row a string-like gas phase is visible along the axis of symmetry but not for $v_{\mathrm{jet}}=1.6\, \mathrm{m/s}$ and $v_{\mathrm{jet}}=1.6\, \mathrm{m/s}$.
Note that this string-like gas phase is an indicator that the planar jet has converged onto the axis of symmetry, as for instance visible along the axis of symmetry in Figure \ref{fig:sim_h1200_p-u}(i).
% since this string-like gas phase could not develop if a regular jet would pierce the bubble from the wall-distant side. 
As the simulations suggest, this is a result of the pinch-off process, i.e., only in the two higher wall jet velocity cases the planar jet achieves to converge on the axis of symmetry, while for the cases $v_{\mathrm{jet}} \leq 1.6\, \mathrm{m/s}$ the regular jet passes the axis already before the planar jet can reach the axis of symmetry.

\subsection{Small stand-off distance}
\begin{figure}
    \centerline{\includegraphics[width=13cm]{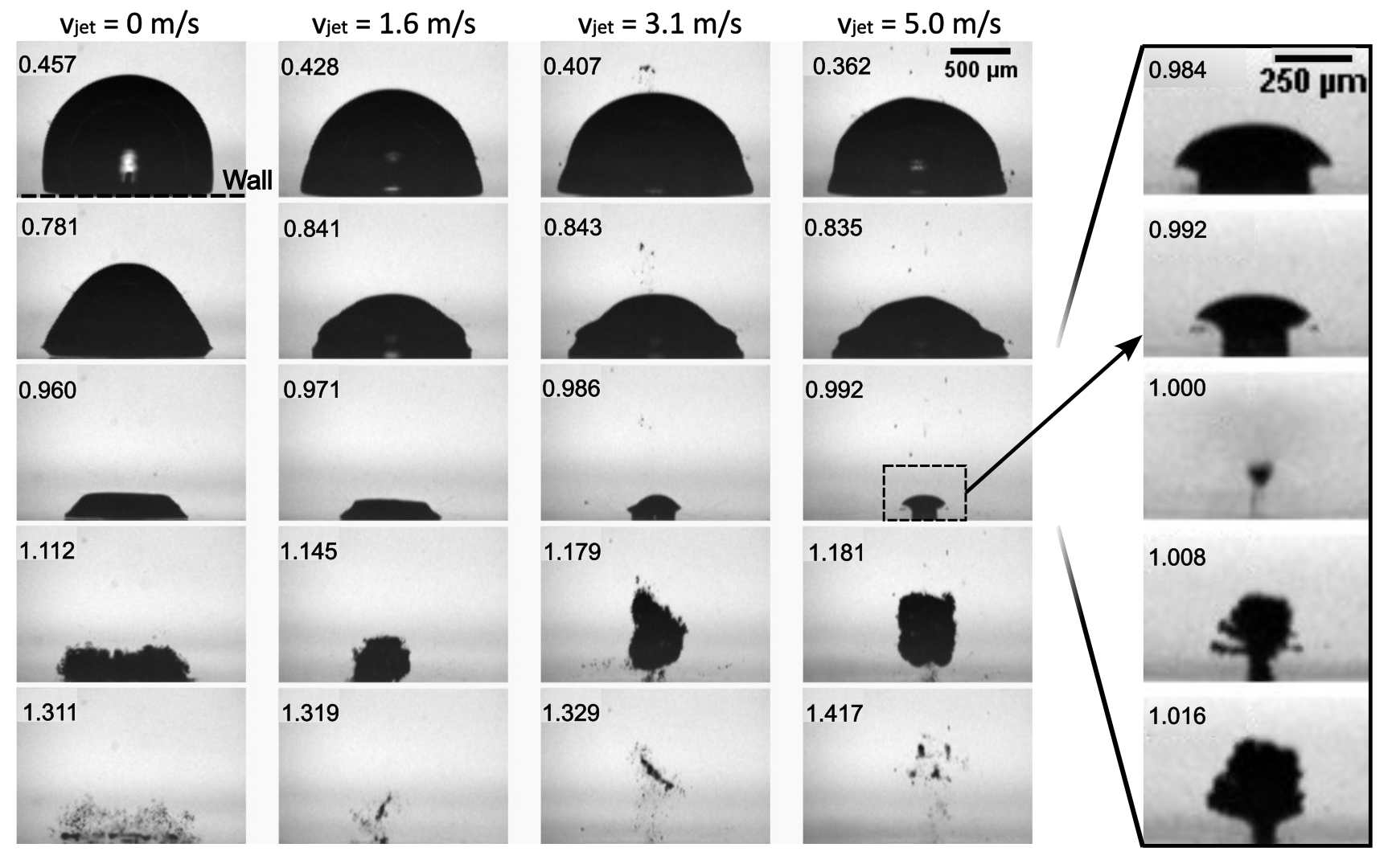}}
\caption{Bubble dynamics at small stand-off ($\gamma=0.49,0.46,0.44,0.46$, from left to right). 
For the fastest wall jet, an upwards-directed jet flow evolves. 
The normalised time is indicated in each frame, the respective lifetimes are $T_\mathrm{L}=151,138,140,127\,\mu$s. The rightmost column shows the collapse for $v_\mathrm{max}$ in detail. The wall is located at the bottom of each frame. A video of the bubble at $v_{jet}=5\,$m/s is provided in the Supplementary Material as Movie 3.
\label{fig: v_Dependence small gamma}}
\end{figure}

\begin{figure}
     \centerline{\includegraphics[width=14cm]{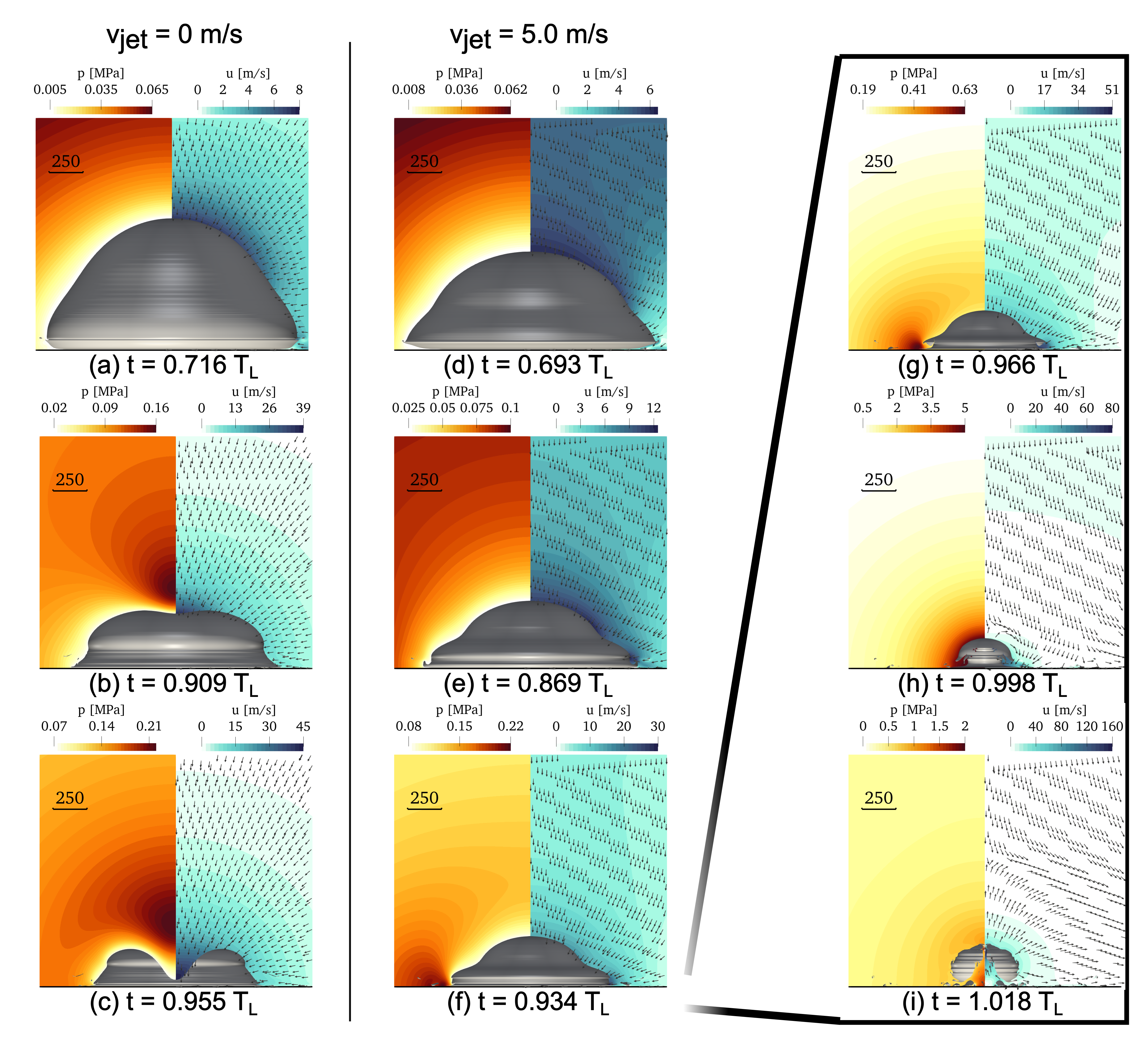}}
    \caption{Bubble shape, pressure and velocity field at different time instances during the collapse of the bubble initially located at $d = 290 \, \mu \mathrm{m}$ ($\gamma = 0.37$ and $0.43$, respectively).
    (a) to (c) In quiescent water ($v_\mathrm{jet} = 0 \, \mathrm{m/s}$). The lifetime of this bubble is $T_\mathrm{L} = 181.6 \, \mu \mathrm{s}$.
    (d) to (i) Subject to a wall jet with $v_\mathrm{jet} = 5 \, \mathrm{m/s}$. The lifetime of this bubble is $T_\mathrm{L} = 155.3 \, \mu \mathrm{s}$. The scale bar corresponds to $250 \, \mu \mathrm{m}$.}
    \label{fig:sim_h0290_p-u}
\end{figure}

\begin{figure}
\centering
    \subfloat[$v_\mathrm{jet} = 0 \, \mathrm{m/s}$, $\gamma = 0.37$]{\includegraphics[width=0.85\linewidth]{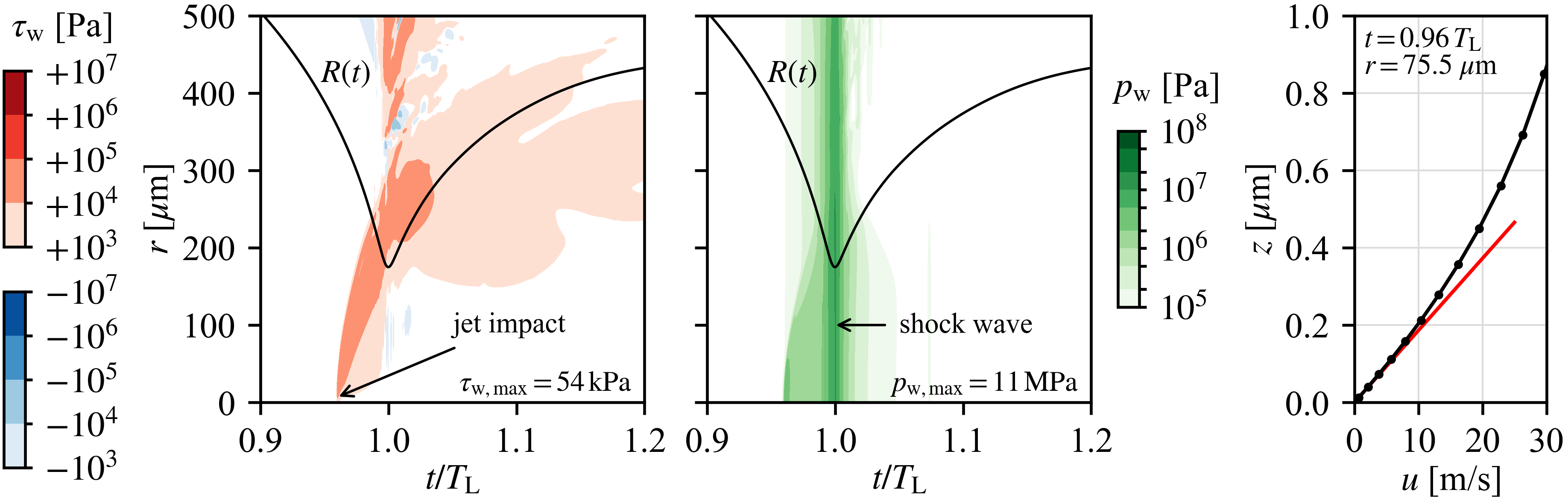}} \\
    \subfloat[$v_\mathrm{jet} = 5 \, \mathrm{m/s}$, $\gamma = 0.43$]{\includegraphics[width=0.85\linewidth]{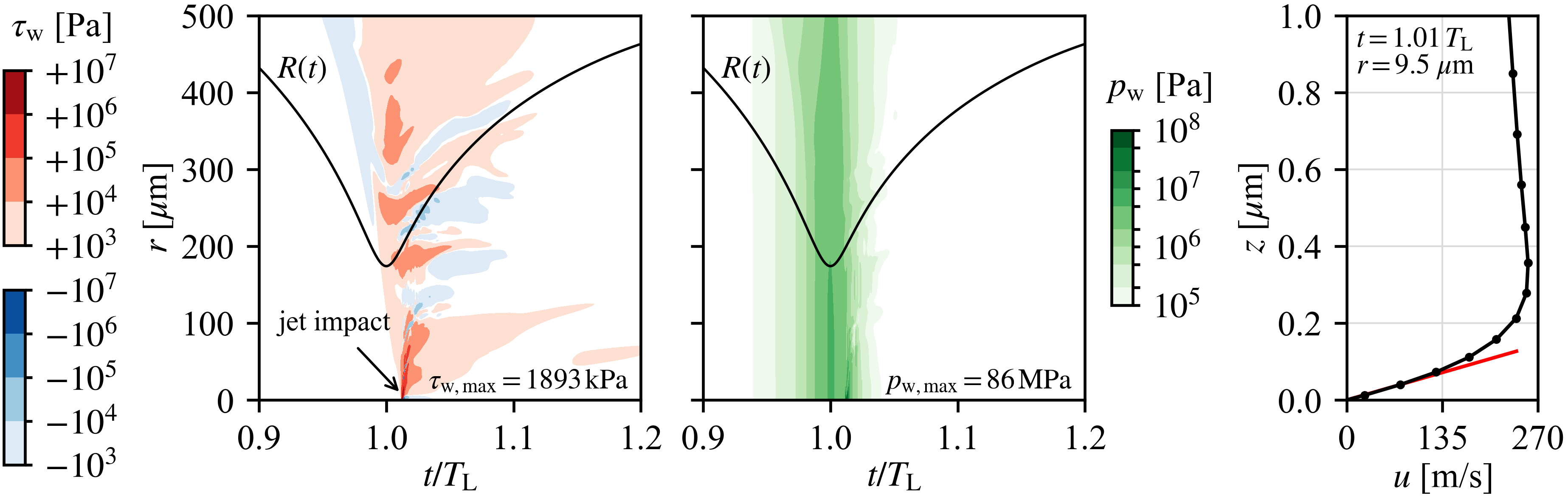}}
    \caption{Space-time plots of the wall shear stress $\tau_\mathrm{w}$ and the wall pressure $p_\mathrm{w}$, and profile of the radial velocity $u$ of the liquid at the location of the highest wall shear rate, of the bubble initially located at $d = 290 \, \mu \mathrm{m}$. In the space-time plots, the black line shows the volume-equivalent bubble radius $R(t)$ and, with respect to $\tau_\mathrm{w}$, red (blue) areas indicate a radially outward (inward) going flow. In the plots of the velocity profiles, the red line represents the velocity gradient associated with $\tau_\mathrm{w,max}$, and the black dots show the locations of the cell centres of the applied computational mesh.}
    \label{fig:bl_h0290}
\end{figure}

The bubble dynamics for the smallest stand-offs presented here ($\gamma\approx 0.45$)
%, with $\gamma$ ranging between 0.44 and 0.49, 
are depicted in Figure \ref{fig: v_Dependence small gamma} for increasing wall jet velocities. The first row shows the maximum expansion, the second row the shape during shrinkage, the third and the fourth row depict the bubble just before and after the collapse, respectively, and the last row shows the bubble at the second collapse.

The quiescent case ($v_{\mathrm{jet}}=0~\mathrm{m/s}$) is similar to the quiescent intermediate $\gamma$ case shown in Figure~\ref{fig: v_Dependence intermediate gamma}. The simulation results are shown in Figure \ref{fig:sim_h0290_p-u}(a)-(c). They reveal a jet velocity of $\approx 45~\mathrm{m/s}$ and a maximum wall shear stress of $54 \, \mathrm{kPa}$ in the space-time plots in Figure \ref{fig:bl_h0290}(a) at $t = 0.964 \, T_\mathrm{L}$, due to the jet impact before the bubble reaches its minimum volume. The maximum pressure peak of $11 \, \mathrm{MPa}$, however, is caused by the collapse of the bubble.

When introducing a wall jet flow with $v_{\mathrm{jet}}=5~\mathrm{m/s}$, in the fourth column of Figure \ref{fig: v_Dependence small gamma}, the bubble during its expansion already moves significantly towards the wall and then shows a quite different dynamics. A video of the corresponding bubble is available as Movie 3 of the Supplementary Material. The bubble extends to a rather flat shape along the wall and a kink is formed during collapse stage, see $t = 0.835 \, T_\mathrm{L}$ and Figure~\ref{fig:sim_h0290_p-u}(e). This kink again is related to the planar and converging jet flow that now develops directly at the wall. This planar jet now converges in very close distance to the wall and proceeds here even below the bubble, such that it gives the bubble a mushroom-like shape, see $t = 0.992 \, T_\mathrm{L}$ in Figure \ref{fig: v_Dependence small gamma} and the magnification in Figure~\ref{fig:sim_h0290_p-u}(h). Upon convergence of the planar jet, the bubble is lifted off the wall and an axial jet is formed, similar as before in the case of a larger stand-off, but this time in the reverse direction, i.e., away from the wall with velocities of $\approx 160 \, \mathrm{m/s}$, see Figure~\ref{fig:sim_h0290_p-u}(i). As a result, the final collapse occurs about $120\,\mu$m away from the wall, see $t = 1.000 \, T_\mathrm{L}$. 
Subsequently, the gas phase, now in the several bubbles arranged along a ring, translates further upward such that the second collapse occurs at a distance of $500\,\mu \mathrm{m}$ from the wall ($t = 1.409\, T_\mathrm{L}$). 

However, even if it is not possible to see directly in the presented graphs, the space-time plot in Figure \ref{fig:bl_h0290} (b) indicates that also a downwards-directed jet hits the wall shortly after the bubble reaches its minimum volume, visible through the wall shear stress peak at $t = 1.01 \, T_\mathrm{L}$. Upon impact, the jet generates a wall shear stress of $1893 \, \mathrm{kPa}$, which is $35$ times larger compared to the stationary case. At the same time, the maximum wall pressure reaches $86 \, \mathrm{MPa}$, which is about $8$ times larger compared to cavitation bubbles in a quiescent liquid and which corresponds to the stagnation pressure of a jet with a velocity of $\approx 410 \, \mathrm{m/s}$.

 %a wall detaching flow is in the quiscient case with similar dynamics produced only for gamma<0.1

Comparing the bubble shapes of the two middle columns just prior and after the collapse in Figure \ref{fig: v_Dependence small gamma}, suggests that in the case of $v_{\mathrm{jet}}=1.6~\mathrm{m/s}$ the regular jet that forms axially pierces the bubble already before the planar boundary-parallel inflow can converge, see $t = 0.992 \, T_\mathrm{L}$. After impact on the wall, the regular jet collides with and inhibits further planar inflow towards the axis of symmetry. In contrast, for $v_{\mathrm{jet}}=3.1~\mathrm{m/s}$ the wall-parallel flow is sufficiently fast to meet on the axis of symmetry before the regular jet pierces the bubble, such that the wall-parallel flow can still lift the bubble off upon convergence. However, the collapse seems to occur closer to the wall than in the case with $v_{\mathrm{jet}}=5.0~\mathrm{m/s}$.

\subsection{Parameter overview}
\label{sec:parameter overview}

\begin{figure}
     \centerline{\includegraphics[width=13cm]{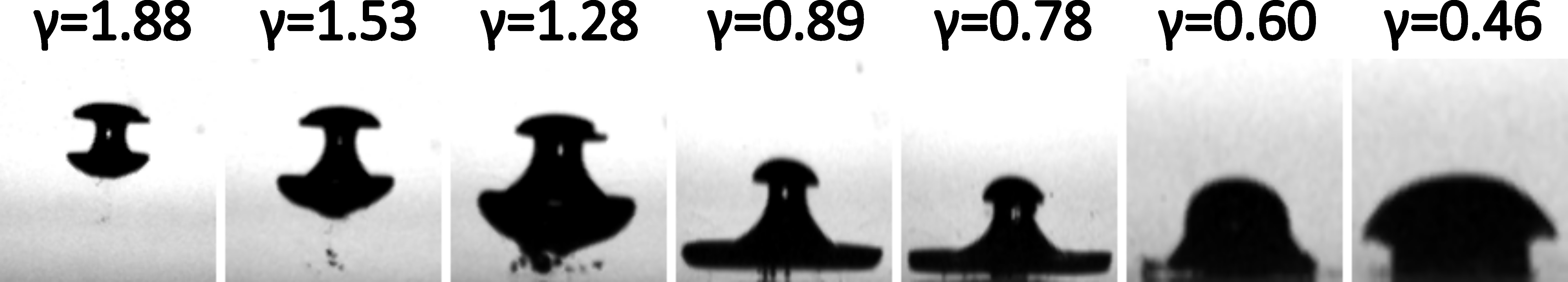}}
    \caption{Bubble shape in the last instance before the collapse for different stand-off distances $\gamma$ (indicated above each frame) with the constant wall jet velocity of $v_\mathrm{jet} = 5 \, \mathrm{m/s}$. The wall coincides with the bottom in all frames. The scale differs in the frames.}
    \label{fig:gamma_change}
\end{figure}

\begin{figure}
    \centerline{\includegraphics[width=13cm]{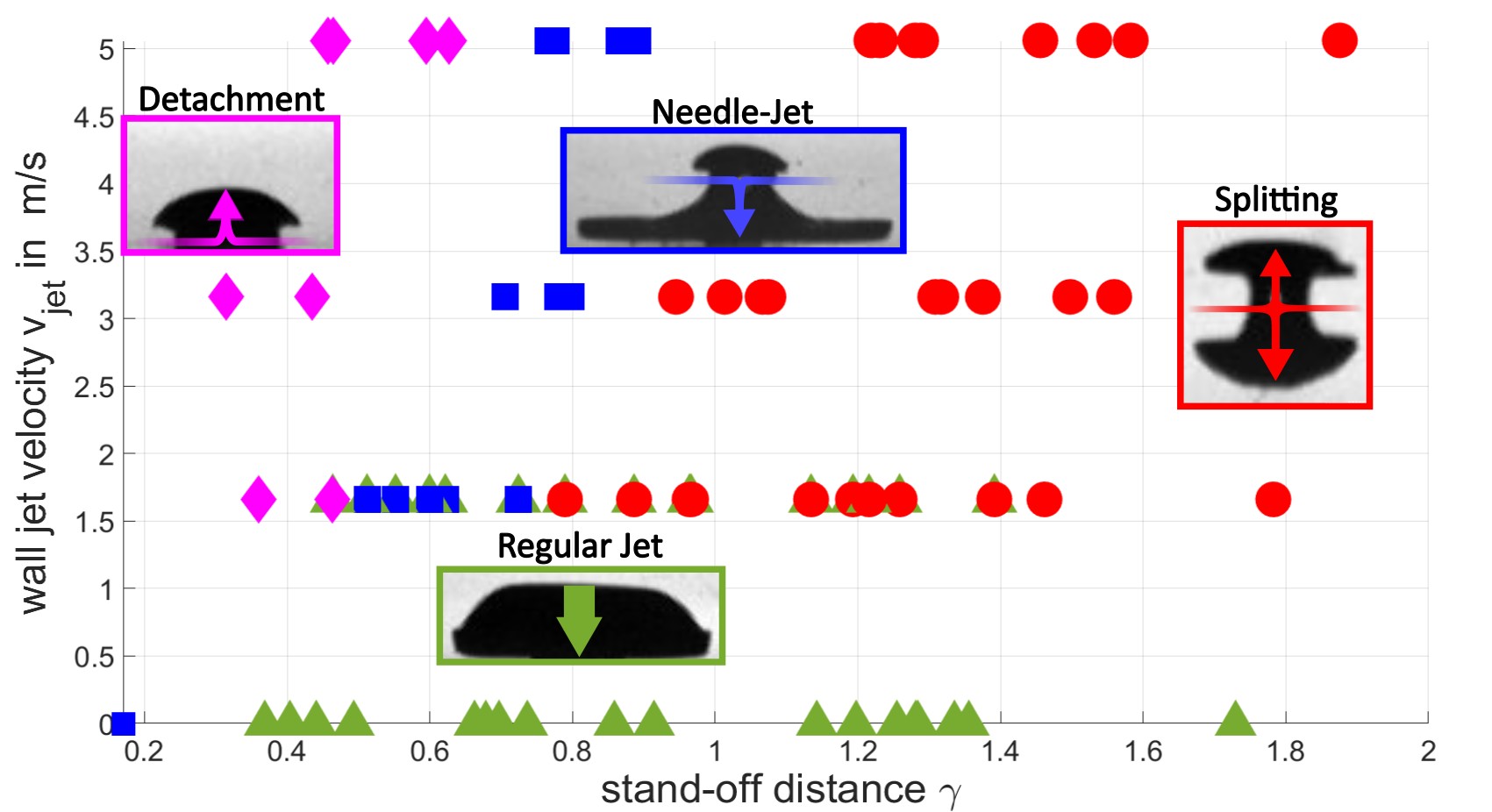}}
    \caption{Parameter overview. Splitting regime (red dots), needle-jet regime (blue squares), detachment regime (magenta diamonds) and regular jetting regime (green triangles) are shown. The arrows on the example pictures indicate the main jetting direction during the collapse. With needle-jet data for $v_{\mathrm{jet}}=0\,$m/s from \cite{Reuter2021}.}
    \label{fig:overview}
\end{figure}

While in the last section, the effect of the wall jet flow on the bubble shape dynamics is studied, in Figure \ref{fig:gamma_change} we compare the stand-off dependence of the bubble shape for the maximum wall jet velocity of  $v_{\mathrm{jet}}=5\,$m/s.
For large stand-offs ($\gamma=1.88$), kink and pinch-off occur rather symmetric, i.e. close to the bubble equator. As the bubble is generated closer to the wall, i.e. for decreasing $\gamma$, the bubble extends further along the wall. There, the bubble dynamics are slowed down and the wall-distal part above the pinch-off region collapses faster, resulting in a small cap being shaped, see for example $\gamma=0.89$. Only at the smallest stand-off ($\gamma=0.46$) shown, the planar jet evolves sufficiently fast and sufficiently close to the wall to result in a bubble lift-off.

In Figure \ref{fig:overview} the jetting behaviour as a function of stand-off and wall jet velocity is presented. In this parameter map, we divide the jetting behaviour into four regimes. First, the regular jet, i.e. the rather broad/thick/wide jet that is directed axially towards the wall, being a few hundred micrometres in diameter and reaching typical velocities of 50-150 m/s (green). Second, the splitting regime, in which the rather symmetric bubble splits upon the convergence of the planar jet (red) with the generation of two oppositely directed jets. Third, the needle-jet regime, i.e. the convergence of the planar jet that results in a much faster jet axially directed toward the wall (blue).
And fourth, the detachment regime, where the planar convergent jet runs at the wall pushing between bubble and wall, converging before the regular jet pierces the bubble, resulting in bubble lift-off (magenta).

Figure \ref{fig:overview} shows how certain regimes extend to larger stand-off values with increasing wall jet velocity. 
For example, the needle-jet regime, in a quiescent liquid only occurring for $\gamma\lessapprox0.2$, extends to larger stand-offs of about $\gamma\approx 0.8$ for $v_{\mathrm{jet}}=5\,$m/s. 
Similarly, the detachment regime is reported in quiescent conditions for $\gamma\approx0.2$, but not $\gamma=0.1$ \citep{abedini2023situ}, and occurs here up to $\gamma\approx 0.6$. 
%With a wall jet this regime can be extended to larger stand-offs ... This is a particularly interesting case because....

Under some conditions both the regular jet and a needle jet or splitting is observed, see for example the $v_{\mathrm{jet}}=1.6\,$m/s row, indicated by the stacked symbols. The corresponding dynamics are shown in Appendix B in more detail.

\begin{figure}
    \centering
    \includegraphics[width=0.85\linewidth]{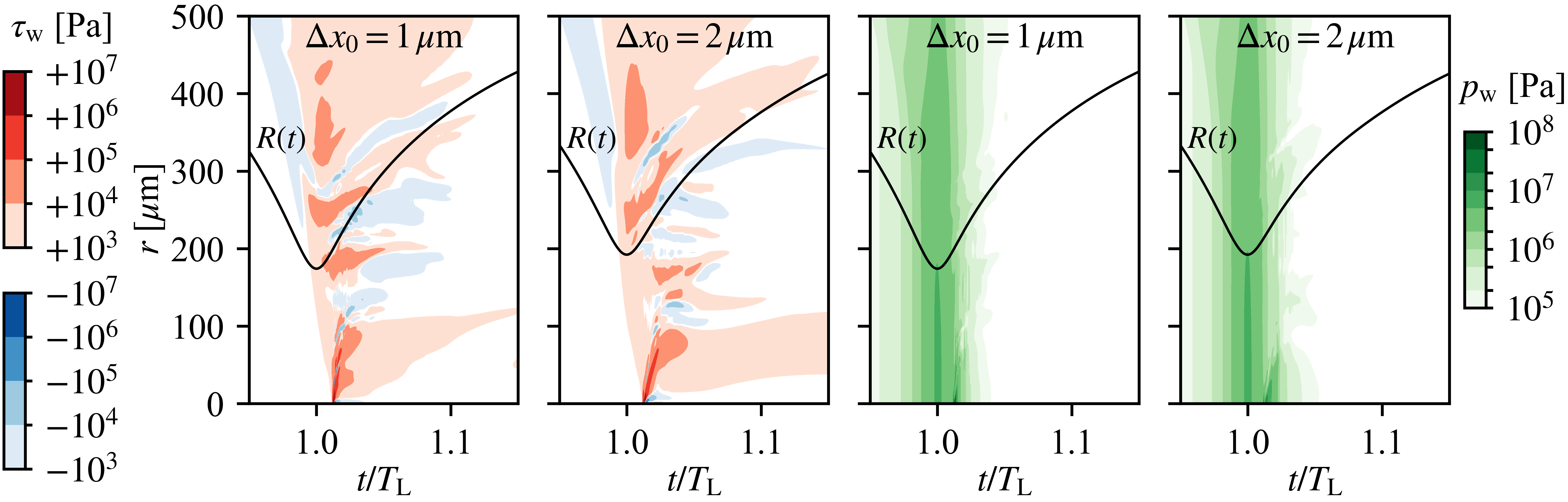}
    \caption{Space-time plots of the wall shear stress $\tau_\mathrm{w}$ and the wall pressure $p_\mathrm{w}$, of the bubble initially located at $d = 290 \, \mu \mathrm{m}$ and subject to wall jet with a velocity of $v_\mathrm{jet}= 5$ m/s, obtained on computational meshes with a mesh spacing of $\Delta x_0 = 1 \, \mu \mathrm{m}$ and $\Delta x_0 = 2 \, \mu \mathrm{m}$. In the space-time plots, the black line shows the volume-equivalent bubble radius $R(t)$ and, with respect to $\tau_\mathrm{w}$, red (blue) areas indicate a radially outward (inward) going flow.}
    \label{fig:bl_h0290_dx}
\end{figure}

\begin{figure}
    \centering
    \includegraphics[width=0.85\linewidth]{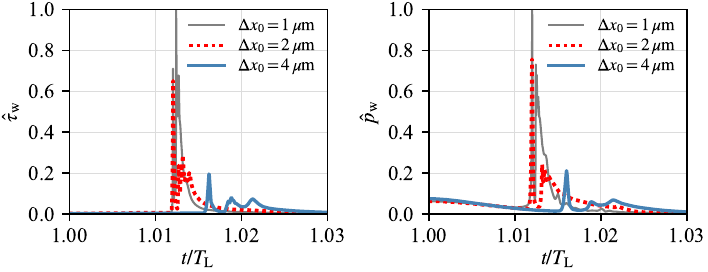}
    \caption{Maximum wall shear stress $\hat{\tau}_\mathrm{w}$ and maximum wall pressure $\hat{p}_\mathrm{w}$, normalized by the respective values on the finest mesh ($\Delta x_0 = 1 \, \mu \mathrm{m}$), of the bubble initially located at $d = 290 \, \mu \mathrm{m}$ and subject to wall jet with a velocity of $v_\mathrm{jet}= 5$ m/s, as a function of time obtained on computational meshes with a mesh spacing of $\Delta x_0 \in \{ 1, 2, 4 \} \, \mu \mathrm{m}$.}
    \label{fig:bubble_h0290_u5_dx}
\end{figure}

\section{Summary and discussion} 
\label{sec:summary and discussion}
Under the assumption of two-dimensional axisymmetry, the maximum wall shear stress and wall pressure recorded in the conducted simulations may be considered as lower bounds of these quantities, since full mesh convergence of these quantities could not be achieved with the available computing resources. Taking the case of the bubble initially located at $d =290 \, \mu \mathrm{m}$ and subject to wall jet with a velocity of $v_\mathrm{jet}= 5$ m/s as an example, the space-time plots of the wall shear stress and wall pressure obtained on meshes with $\Delta x_0 \in \{1,2\} \, \mu \mathrm{m}$, shown in Figure \ref{fig:bl_h0290_dx}, exhibit very little qualitative differences. The corresponding maximum values for the wall shear stress and wall pressure obtained on meshes with $\Delta x_0 \in \{ 1, 2, 4 \} \, \mu \mathrm{m}$ are shown in Figure \ref{fig:bubble_h0290_u5_dx}. A converging trend towards higher peak values can be observed as the mesh is refined. Furthermore, the time at which the peak values occur exhibits a negligible difference between the results obtained on meshes with $\Delta x_0 =1 \, \mu \mathrm{m}$ and  $\Delta x_0 = 2 \, \mu \mathrm{m}$.
This is consistent with previous studies regarding the cavitation-generated wall pressure \citep{Mihatsch2015, Trummler2021}, which demonstrated that the maximum wall pressure is strongly mesh dependent, with increasing peak values for decreasing mesh spacing.
Note that the centres of the mesh cells closest to the wall are located at a distance of only $12.5 \, \mathrm{nm}$  ($\Delta x_0 = 1 \, \mu \mathrm{m}$) and $25 \, \mathrm{nm}$  ($\Delta x_0 = 2 \, \mu \mathrm{m}$), highlighting the extreme resolution requirements of these simulations.
However, the maximum jet velocity generated in experimental settings is limited by the symmetry of the needle jet generation \citep{gordillo2023jets} and the jet stability \citep{reuter2022cavitation}. Both effects cannot be covered in axisymmetric simulations.

In cavitation-solid interaction jetting is a key aspect as it shapes the bubble dynamics and due to its direct interaction with the wall. The pressure gradient of a stagnation point flow alters the jetting behaviour and the bubble dynamics significantly. The bubble translates with the flow towards the wall while the pressure gradient caused by the stagnation flow tends to push the bubble away from the wall \citep{Blake2015}. As a consequence of the opposing pressure gradients, the bubble takes an oblate shape which is decisive for the further bubble dynamics.
During collapse, the oblate bubble develops a planar converging jet. This planar jet shapes the initially oblate bubble to an hourglass shape and can even result in bubble pinch off.
These dynamics were studied by \citet{Starrett1982} and in numerical studies by Blake and co-workers \citep{blake_taib_doherty_1986,robinsonblake1994,Blake2015} for rather large bubble-to-wall stand-offs and is a general feature of the collapse of an oblate-spheroidally shaped bubble even without stagnation flow \citep{Chapman1972,Voinov1975,shima1981collapse}.

When the planar jet converges onto the slender gas volume on the axis of symmetry, a high-pressure stagnation point occurs, the gas phase collapses and a thin needle jet is ejected along the axis of symmetry \citep{gordillo2023jets}. This type of jetting has also been shown to occur in stagnant liquid, but only for the smallest bubble to wall stand-offs of about $\gamma \lessapprox 0.2$ \citep{Lechner2019,Reuter2021,Bussmann2023}. This regime is extended to larger stand-offs with the stagnation flow. 

The needle jet at larger stand-offs stimulated by the stagnation flow allows achieving about $40$ times higher wall shear at only around $5$ times larger wall pressures. The simulations show, that at the wall the jet impact pressures in the stagnation flow case are larger than the bubble collapse pressures, which stands in contrast to the bubble dynamics in stagnant flow, where only regular jetting occurs, and the collapse pressures are larger than jet impact pressures.

Recently, it has been shown that only the bubble collapse directly at the wall is erosive, which occurs for $\gamma\lessapprox 0.2$ for a stagnant liquid \citep{reuter2022cavitation}. It is fascinating to observe that the bubble in the small $\gamma$ regime does not collapse at the wall, but lifts off and collapses $\approx 120\,\mu$m away in the bulk, and indeed, we did not find indications of material damage on our glass samples here. Consistently, the maximum wall shear stresses, and wall pressures found here over the entire range of stand-off distances stay mainly below the yield strength and tensile strength of most metals or metal alloys, such as silver with $55 \, \mathrm{MPa}$ and $110-340 \, \mathrm{MPa}$, respectively \citep{smith1995low} and stainless steel 316 L with $440 \, \mathrm{MPa}$ and  $1020 \, \mathrm{MPa}$, respectively \citep{Franc2009}. %Only the lift-off bubble at the smallest stand-off distance reaches a sufficiently large peak wall pressure of $86 \, \mathrm{MPa}$ to potentially cause any damage on silver at all.

%The considered parameter space shows that the $\gamma$-range for the different collapse regimes shifts toward larger stand-off distances for increasing wall jet flow velocities, which is particularly prominent for the needle-jet regime when we take into account the data from \cite{Reuter2021} for $v_{\mathrm{jet}}=0\,$m/s. This phenomenon may be exploited in practice since bubbles with larger stand-off distances are technically easier to realise. 

The adjustment of the wall jet flow velocity and the stand-off distance allows us to control the behaviour of the bubble precisely and, thus, achieve a tailored interaction with the wall. Due to the rich bubble dynamics, a wide range of applications is possible. One example is a cavitation bubble at a small stand-off distance to the wall in a wall jet flow with sufficient velocity. Here, high wall shear stress can be achieved when the jet hits the wall, which can be used for precision cleaning. At the same time, the bubble detaches from the wall and thus is not in contact with the wall during the collapse. We anticipate that this could avoid material damage. 

%****im symmetrischen Fall ist das Material sicher vor dem Jet dann im unsymmetrischen umso mehr

\section{Conclusion}
We studied the dynamics of a single cavitation bubble close to a wall in the stagnation flow of a wall jet. Using high-speed imaging together with numerical simulations, which provide quantitative insight into the produced wall shear stresses and wall pressures, we found that a wall jet flow of already rather small velocities, i.e., of less than $10\,\mathrm{m/s}$, can shape the bubble and significantly change its dynamics. As a result, wall shear stresses and pressures exerted onto the nearby wall are altered drastically.
For example, the wall shear stresses, which are crucial in surface cleaning, can be increased by a factor of 44 as compared to the stagnant case and reach values above $2700\,$kPa here. At the same time, the wall pressures, which are considered a prime cause for cavitation erosion, are increased by only a factor of six as compared to the stagnant case and reach values of $49\,$MPa, i.e. stay below typical material damage thresholds. 

As cavitation bubbles in a stagnant liquid have been shown to be erosive only when they collapse directly at the wall, it is an important finding that a cavitation bubble in the wall jet, somewhat counterintuitively, can lift off the wall just before its collapse and migrate against the wall jet flow direction such that its collapse occurs without wall contact.

The mechanisms of how the wall jet alters the bubble dynamics can be understood via the oblate, ellipsoidal shaping of the bubble by the wall jet. This stimulates the formation of wall-parallel and convergent planar jets. They are also the origin of needle jet formation and consequently, the needle jet regime, in stagnant liquid only found for bubbles at very small stand-offs, extends further into larger stand-off ranges with increasing wall jet velocities.

As the wall jet is technologically rather easy to implement, our results suggest it can be exploited to tailor bubble-wall interaction in terms of stresses on the wall and their distribution, with potential applications for example in material peening and hardening, surface cleaning and erosion prevention.
%Still, further studies are needed to explore this large parameter space of gamma and wall jet velocity $v_\mathrm{jet}$ where the desired outcome depends on the application, either damage needs to be avoided or large wall shear achieved. Additionally, erosion studies with various surface materials can reveal the actual damage potential of the discussed bubble regimes.

%\begin{supplementary}
%Movie 1: High-speed movie of the bubble dynamics of Figure \ref{fig: v_Dependence large gamma} recorded at 1 Million frames per second ($\gamma=1.88$, $v_\mathrm{jet}= 5$ m/s). A waist forms at the bubble equator and pinches off with the formation of two rather thin and fast jets in opposite directions.

%Movie 2: High-speed movie of the bubble dynamics of Figure \ref{fig: v_Dependence intermediate gamma} recorded at 1 Million frames per second ($\gamma=0.78$, $v_\mathrm{jet}= 5$ m/s). The bubble pinches off with the bottom flat above the wall and a high-speed needle-like jet towards the wall forms.

%Movie 3: High-speed movie of the bubble dynamics of Figure \ref{fig: v_Dependence small gamma} recorded at 1 Million frames per second ($\gamma=0.46$, $v_\mathrm{jet}= 5$ m/s). The planar jet occurs above the wall and forms a mushroom-shaped bubble that lifts off the wall prior to the collapse.

%Movie 4: High-speed movie of the bubble dynamics of Figure \ref{fig: transitional bubble} recorded at 1 Million frames per second ($\gamma=0.96$, $v_\mathrm{jet}= 1.6$ m/s). A regular jet flow and a planar jet flow occur simultaneously.
%\end{supplementary}

~\\
\textbf{Funding.} We thank the Alfred Kärcher Foundation and the DFG through INST 272/280-1 for their financial support.

~\\
\textbf{Declaration of Interests.} The authors report no conflict of interest.

\appendix 
\section{Flow field}
\label{sec: Appendix A}

\begin{figure}
\centerline{\includegraphics[width=7cm]{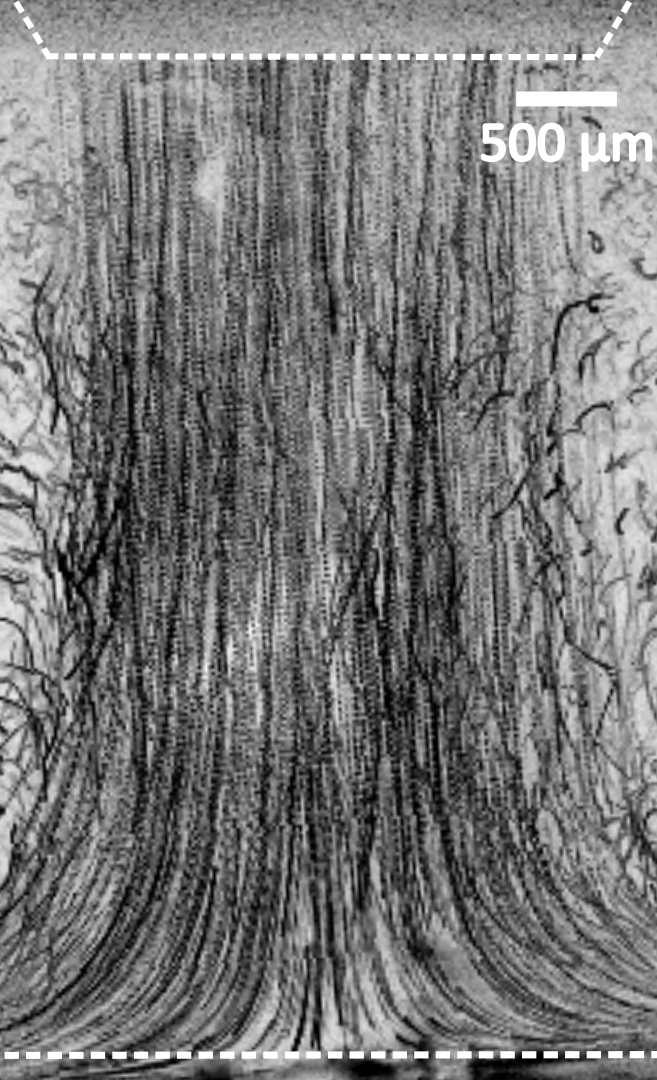}}
\caption{Side view showing the pathlines obtained from microscopic particles seeded into the wall jet flow and recorded with a high-speed camera. The top lines indicate the nozzle exit and the bottom line the position of the wall, $v_{\mathrm{jet}}=5~\mathrm{m/s}$.}
\label{fig: flow profile}
\end{figure}

To observe the flow profile of the wall jet, we add particles with about $20\,\mu$m in diameter to the liquid and record their motion with a high-speed camera. 80 images recorded with a frame rate of 200,000 are summed up to obtain pathlines of the nozzle flow, as shown in Figure \ref{fig: flow profile} for a flow velocity of $v_{\mathrm{jet}}=5~\mathrm{m/s}$. At the top, the flow leaves the nozzle exit, and it impinges onto the wall at the bottom of Figure \ref{fig: flow profile}. Then the flow spreads radially outwards, yet it remains mostly laminar without much of a recirculation or disturbance.

\section{Transition regimes}
\label{sec:Appendix B}

\begin{figure}
\centerline{\includegraphics[width=13cm]{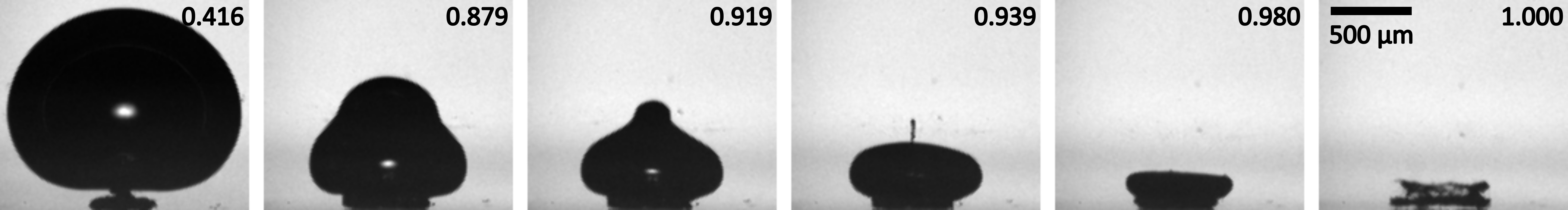}}
\caption{Bubble shape at $\gamma=0.96$ and $v_\mathrm{jet}= 1.6$ m/s with both regular jet and planar jet resulting in pinch-off. The normalised time is indicated in each frame, the bubble lifetime is $T_\mathrm{L}=149\,\mu$s. A movie of this series is provided as Movie 4 in the Supplementary Material.}
\label{fig: transitional bubble}
\end{figure}

An example of a bubble that shows pinch-off from the planar jet flow and that gets also pierced by the regular jet is presented in Figure \ref{fig: transitional bubble} for a bubble at $\gamma=0.96$. 
Between $t=0.919$ and $0.939$, the planar flow meets at the axis of symmetry and results in the collapsed string-like gas phase along the axis of symmetry. However, probably due to the limited momentum associated with the planar flow, at later times the regular jet pierces the bubble and results in the torus collapse geometry known from quiescent liquid.

\end{document}